\documentclass[twocolumn]{aastex631}
\usepackage{CJK}
\usepackage{graphicx}
\usepackage{amsmath}	% Advanced maths commands
\usepackage{amssymb}
\usepackage[figuresleft]{rotating}

\submitjournal{AAS Journal}

\shorttitle{Properties of dust-obscured QSOs}
\shortauthors{Sun et al.}

\begin{document}
\begin{CJK*}{UTF8}{gbsn}

\title{Physical properties of hyperluminous, dust-obscured quasars at $z \sim 3$: multiwavelength Spectral Energy Distribution analysis and cold gas content revealed by ALMA}

\correspondingauthor{Lulu Fan}
\email{llfan@ustc.edu.cn}

\author[0009-0004-7885-5882]{Weibin Sun (孙卫斌)}
\affiliation{CAS Key Laboratory for Research in Galaxies and Cosmology, Department of Astronomy,
University of Science and Technology of China, Hefei 230026, China}
\affiliation{School of Astronomy and Space Science, University of Science and Technology of China, Hefei 230026, China}

\author[0000-0003-4200-4432]{Lulu Fan (范璐璐)}
\affiliation{CAS Key Laboratory for Research in Galaxies and Cosmology, Department of Astronomy,
University of Science and Technology of China, Hefei 230026, China}
\affiliation{School of Astronomy and Space Science, University of Science and Technology of China, Hefei 230026, China}
\affiliation{Deep Space Exploration Laboratory, Hefei 230088, China}

\author[0000-0002-2547-0434]{Yunkun Han (韩云坤)} 
\affiliation{Yunnan Observatories, Chinese Academy of Sciences, 396 Yangfangwang, Guandu District, Kunming, 650216, P. R. China}
\affiliation{Center for Astronomical Mega-Science, Chinese Academy of Sciences, 20A Datun Road, Chaoyang District, Beijing, 100012, P. R. China}
\affiliation{Key Laboratory for the Structure and Evolution of Celestial Objects, Chinese Academy of Sciences, 396 Yangfangwang, Guandu District, Kunming, 650216, P. R. China}

\author[0000-0002-7821-8873]{Kirsten K. Knudsen}
\affiliation{Department of Space, Earth and Environment, Chalmers University of Technology, Onsala Space Observatory, SE-439 92 Onsala, Sweden}

\author[0000-0002-4742-8800]{Guangwen Chen （陈广文）}
\affiliation{CAS Key Laboratory for Research in Galaxies and Cosmology, Department of Astronomy, University of Science and Technology of China, Hefei 230026, China}
\affiliation{School of Astronomy and Space Science, University of Science and Technology of China, Hefei 230026, China}

\author[0000-0003-1632-2541]{Hong-Xin Zhang (张红欣)}
\affiliation{CAS Key Laboratory for Research in Galaxies and Cosmology, Department of Astronomy, University of Science and Technology of China, Hefei 230026, China}
\affiliation{School of Astronomy and Space Science, University of Science and Technology of China, Hefei 230026, China}

\begin{abstract}

We present a UV to millimeter spectral energy distribution (SED) analysis of 16 hyperluminous, dust-obscured quasars at z $\sim$ 3, selected by the \textit{Wide-field Infrared Survey Explorer}. We aim to investigate the physical properties of these quasars, with a focus on their molecular gas content. We decompose the SEDs into three components: stellar, cold dust, and active galactic
nucleus (AGN). By doing so, we are able to derive and analyze the relevant properties of each component. We determine the molecular gas mass from CO line emission based on Atacama Large Millimeter/submillimeter Array (ALMA) observations. By including ALMA observations in the multiwavelength SED analysis, we derive the molecular gas fractions, gas depletion timescales, and star formation efficiencies (SFEs). Their sample median and 16th-84th quartile ranges are $f_{\rm gas}\,\sim\,0.33_{-0.17}^{+0.33}$, $t_{\rm depl}\,\sim$ 39$_{-28}^{+85}$ Myr, SFE $\sim\,$ 297$_{-195}^{+659}$ $\rm K\,\rm km\,\rm s^{-1}\,\rm pc^{-2}$. Compared to main-sequence galaxies, they have a lower molecular gas content and higher SFEs, similar to quasars in the literature. This suggests that the gas in these quasars is rapidly depleted, likely as the result of intense starburst activity and AGN feedback. The observed correlations between these properties and the AGN luminosities further support this scenario. Additionally, we infer the black hole to stellar mass ratio and black hole mass growth rate, which indicate a significant central black hole mass assembly over short timescales. Our results are consistent with the scenario that our sample represents a short transition phase toward unobscured quasars.

\end{abstract}


\section{Introduction} \label{sec:intro}

The coevolution between the central supermassive black hole (SMBH) and host galaxy is now widely acknowledged \citep{{Kormendy2013}}. This is evidenced by the tight correlation between the mass of central SMBHs and the stellar bulge masses in galaxies \citep[e.g.,][]{Magorrian1998, Ferrarese2005}. In one of the most popular coevolution scenarios, galaxy gas-rich major galaxy mergers trigger intense starbursts, provide the fuel for central SMBH accretion, and trigger active galactic nucleus (AGN) activity delayed after the triggering of starbursts \citep{Hopkins2008}. This stage of AGN-starburst composite systems often leads to the observation of galaxies as dust-obscured quasars. The galaxies will evolve to unobscured quasars after the accreting SMBH experienced a “feedback” phase, which clears the dust and gas in the galaxy in the form of powerful outflowing winds \citep[see][for recent reviews]{fabian2012,Somerville2015}. During this evolutionary sequence first presented by \citet{Sanders1988}, dust-obscured quasars have been identified as essential links between starbursts and unobscured quasars. They serve a critical role in the rapid assembly of both the SMBH and galaxy mass, as well as in AGN feedback \citep{Hickox2018}. Evidence from both observations and theoretical models have suggested that the efficiency of galaxy-scale outflows increases with quasar bolometric luminosity \citep[see, e.g.,][]{Heckman2014,Hopkins2016,Fiore2017,Garca2021}. Therefore, luminous, obscured quasars are good candidates for investigating the interplay between host galaxies and their central SMBHs.

The \textit{Wide-field Infrared Survey Explorer} \citep[WISE;][]{Wright2010} has revealed an important population of luminous, dust-obscured galaxies at z $\sim$ 3 by selecting sources that are strongly detected at 12 and 22 $\mu$m, but weakly or not detected at 3.4 and 4.6 $\mu$m \citep{Eisenhardt2012}. A variety of follow-up studies utilizing different techniques have been carried out. Multiband spectral energy distribution (SED) analyses have played an essential role in the study of these high-redshift objects. Through the construction of median IR SEDs, it has been revealed that these galaxies exhibit a high mid-infrared (MIR) to submillimeter luminosity ratio, elevated dust temperatures, and extraordinary bolometric luminosities over $10^{13}\,L_{\odot}$ \citep{Wu2012}. The number density of these luminous, DOGs is comparable to that of equally luminous type 1 quasars \citep{Assef2015}. These galaxies represent an exceedingly rare class of DOGs, now commonly referred to as hot, dust-obscured Galaxies (Hot DOGs). Further investigations have shown that the IR SEDs of the most luminous Hot DOGs are dominated by hot dust at a temperature exceeding 450 K \citep{Tsai2015}, and their IR SEDs can be decomposed through two component, AGN-starburst SED fitting \citep{Fan2016b, Fan2017a,Fan2018a}. UV-optical spectral analyses of Hot DOGs found black hole masses $\sim\,10^{9}\,M_\odot$ which are accreting near or above the Eddington limit and also host powerful ionized outflows \citep{Wu2012,  Tsai2018, Wu2018, Finnerty2020, Jun2020}. Millimeter interferometric observations like Atacama Large Millimeter/submillimeter Array (ALMA) of several Hot DOGs revealed a highly turbulent ISM and also provide evidence of possible molecular outflows \citep{Wu2014,Diaz-santos2016,Diaz-santos2018,Diaz-santos2021,Fan2018b,Fan2019,Penney2020}. The X-ray observations of Hot DOGs consistently find high column densities close to Compton thick \citep{Stern2014,Piconcelli2015,Assef2016,Assef2020,Ricci2017,Vito2018,Zappacosta2018}. The environments where Hot DOGs reside were found to be significantly overdense \citep{Jones2015,Jones2017,Silva2015,Penney2019,Ginolfi2022,Luo2022,Zewdie2023}. All results are generally consistent with the merger-driven coevolution scenario.

SED fitting is an effective method for decomposing emissions between star formation and AGN \citep{Sokol2023}. Some widely used SED fitting codes, including SED3FIT \citep{Berta2013}, CIGALE \citep{Boquien2019,Yang2022}, and BayeSED \citep{Han2012a,HanY2014a,HanY2019a}, have been proven to be efficient tools for exploring AGN-starburst systems, and AGN models are constantly evolving and becoming more and more accurate and realistic \citep{Fritz2006,Nenkova2008a,Nenkova2008b,Honig2010,Honig2017,Siebenmorgen2015,Stalevski2016}. At high redshift, it is hard to resolve the central AGN emission from the host galaxy. SED observations, modeling, and fitting are indispensable to investigate the physical properties of these high-redshift AGNs and their host galaxies \citep[e.g.,][]{Merloni2010,Bongiorno2014,Suh2019,Lopez2023}.

Molecular gas, predominantly traced by carbon monoxide (CO) emission lines, acts as the fuel for both star formation and black hole accretion. Additionally, it plays an important role in energy feedback from AGN \citep[e.g.,][]{Feruglio2017,Bischetti2019,Fluetsch2019}. Investigations of the molecular gas content ($M_{\rm H_2}$) in quasars, coupled with other observables such as SED-derived dust masses ($M_{\rm dust}$), stellar masses ($M_\star$), and star formation rates (SFRs), provide valuable insights into the physical processes driving the coevolution of galaxies and SMBHs \citep[e.g.,][]{Brusa2015,Brusa2018,Banerji2017,Kakkad2017,Perna2018,Bischetti2021}. However, for Cosmic noon (z $\sim$ 2-3) when both star formation and black hole accretion activity in the Universe peaked \citep{Shapley2011}, there is a lack of systematic investigation into the molecular gas properties of luminous quasars, primarily limited to analyses of individual sources or relatively small sample sizes.

To investigate systematically the physical properties of Hot DOGs at z $\sim$ 3, particularly their cold gas content, we conduct a comprehensive UV to millimeter SED analysis of a subsample of 16 Hot DOGs selected from \citet{Fan2016b}. Table \ref{tab:sample} lists the properties of these Hot DOGs, as reported in \citet{Fan2016b}. This subsample has ALMA observations of CO emission lines, which were used to constrain their molecular gas content. This study represents the largest sample of high-redshift, luminous obscured quasars so far in this kind of study. We derived the properties of stellar, cold dust, AGN, and gas components and calculated their relative values (e.g., molecular gas fraction $f_{\rm gas}$ = $M_{\rm H_2}/(M_{\rm H_2}\,+\,M_\star)$ and SFE), further testing the role of Hot DOGs in massive galaxy and SMBH coevolution. In Section \ref{sec:alma}, we present details of ALMA observations and the subsequent data analysis. Section \ref{section3} covers the construction of our multiwavelength SED and the SED modeling approach we used. Results and discussions are described in Section \ref{section4} and Section \ref{section5}, respectively. Finally, in Section \ref{section6}, we summarize our findings and draw a conclusion. Throughout this work, we assume a Lambda cold dark matter ($\Lambda$CDM) cosmology \citep[see][]{Komatsu2011} with $H_0 = 70$\,km\,s$^{-1}$\,Mpc$^{-1}$, $\Omega_{\rm M} = 0.3$, and $\Omega_\Lambda = 0.7$.  

\begin{deluxetable}{lcccccc}
\tablecaption{Sample properties  \label{tab:sample}}
\tabletypesize{\scriptsize}
\setlength{\tabcolsep}{2pt}
\tablehead{
\\
\colhead{Source} & 
\colhead{R.A.$_{\it WISE}$} & 
\colhead{Dec$_{\it WISE}$} & 
\colhead{Redshift} & 
\colhead{log $L^{\rm t}_{\rm IR}$} & 
\colhead{log $L^{\rm cd}_{\rm IR}$} \\
\colhead{Name} & 
\colhead{(J2000)}  & 
\colhead{(J2000)} & 
\colhead{ } & 
\colhead{log($L_\odot$)} & 
\colhead{log($L_\odot$)}\\  
%\colhead{} 
\colhead{(1)} & \colhead{(2)} & \colhead{(3)} & \colhead{(4)} & \colhead{(5)} & \colhead{(6)}
}
\startdata
W0126$-$0529$^a$ & 01:26:11.96 & $-$05:29:09.6 & 2.937 &	13.95 $\pm$ 0.01 & 13.90 $\pm$ 0.02   \\ 
W0134$-$2922 & 01:34:35.71 & $-$29:22:45.4 & 3.047 &	13.97 $\pm$ 0.03 & 13.15 $\pm$ 0.03   \\
W0149+2350 & 01:49:46.16 & +23:50:14.6 & 3.228 &	13.90 $\pm$ 0.04 & 13.11 $\pm$ 0.07   \\
W0220+0137 & 02:20:52.12 & +01:37:11.6 & 3.122 &	14.07 $\pm$ 0.02 & 13.63 $\pm$ 0.04   \\
W0248+2705 & 02:48:58.81 & +27:05:29.8 & 2.210 &	13.45 $\pm$ 0.05 & 13.11 $\pm$ 0.07   \\
W0410$-$0913 & 04:10:10.60 & $-$09:13:05.2 & 3.592 &	14.12 $\pm$ 0.03 & 13.84 $\pm$ 0.03   \\ 
W0533$-$3401 & 05:33:58.44 & $-$34:01:34.5 & 2.904 &	13.94 $\pm$ 0.04 & 13.50 $\pm$ 0.04   \\
W0615$-$5716 & 06:15:11.07 & $-$57:16:14.6 & 3.399 &	14.13 $\pm$ 0.02 & 13.33 $\pm$ 0.14   \\ 
W1248$-$2154 & 12:48:15.21 & $-$21:54:20.4 & 3.318 &	14.15 $\pm$ 0.02 & 13.18 $\pm$ 0.07   \\ 
W1603+2745 & 16:03:57.39 & +27:45:53.3 & 2.633 &	13.61 $\pm$ 0.02 & 13.32 $\pm$ 0.04   \\ 
W1814+3412 & 18:14:17.30 & +34:12:25.0 & 2.452 &	13.72 $\pm$ 0.03 & 13.12 $\pm$ 0.07   \\
W2054+0207 & 20:54:25.69 & +02:07:11.0 & 2.520 &	13.66 $\pm$ 0.05 & 13.16 $\pm$ 0.06   \\ 
W2201+0226 & 22:01:23.39 & +02:26:21.8 & 2.877 &	13.84 $\pm$ 0.03 & 13.73 $\pm$ 0.02   \\
W2210$-$3507 & 22:10:11.87 & $-$35:07:20.0 & 2.814 &	13.93 $\pm$ 0.02 & 13.47 $\pm$ 0.02   \\ 
W2238+2653 & 22:38:10.20 & +26:53:19.8 & 2.405 &	13.79 $\pm$ 0.03 & 13.48 $\pm$ 0.04   \\ 
W2246$-$0526 & 22:46:07.57 & −05:26:35.0 & 4.593 &    14.46 $\pm$ 0.02 & 13.73 $\pm$ 0.04   \\
W2305$-$0039 & 23:05:25.88 & $-$00:39:25.7 & 3.106 &	13.97 $\pm$ 0.02 & 13.61 $\pm$ 0.03 \\  
\enddata
\tablecomments{
(1): Source names. $^a$ W0126$-$0529 has been excluded from our sample for its ambiguous redshift identification.
(2) and (3): The WISE coordinates from the AllWISE database.
(4): The spectroscopic redshift from \citet{Wu2012} and \citet{Tsai2015}.
(5) and (6): The IR luminosities of AGN torus and cold dust emission derived by IR decomposition as reported in \citet{Fan2016b}.
}
\end{deluxetable}

%----------------------- Table of ALMA observations
\begin{deluxetable*}{llccccl}
\tablecaption{Summary of the ALMA Observations \label{tab:obssum} }
\tabletypesize{\scriptsize}
\tablehead{
\\
\colhead{Source} & \colhead{Line} & \colhead{Date} & \colhead{Flux \& Bandpass} & \colhead{Gain} & \colhead{$\sigma$ $^a$} & \colhead{Beam} \\
\colhead{Name} & \colhead{}  & \colhead{DD-MM-YYYY} & \colhead{Calibrator} & \colhead{Calibrator} & \colhead{(mJy/beam)}  &  \colhead{($'' \times ''$)} 
}
\startdata
W0126$-$0529 &  CO(3-2) & 02-01-2018 & J0006$-$0623 & J0141$-$0202 & 0.45 & $0.61\times0.52$, 82.4$^\circ$ \\ [1pt]
W0134$-$2922 $^b$ & CO(4-3) & 12-12-2017 & J2357$-$5311 & J0120$-$2701 & 0.11 & $0.39\times0.27$, $-63.1^\circ$ \\
               &         & 13-12-2017 & J2357$-$5311 & J0120$-$2701 &  \\ [2pt]
W0248+2705 & CO(4-3) & 10-01-2018 & J0238+1636 & J0237+2848 & 0.42 & $0.58\times0.56$, 23.9$^\circ$ \\
W0533$-$3401 &  CO(3-2) & 20-12-2017 & J0538$-$4405 & J0522$-$3627 & 0.32 & $0.60\times0.51$, $-78.9^\circ$ \\
W0615$-$5716 $^b$ & CO(4-3) & 07-12-2017 & J0519$-$4546 & J0550$-$5732 & 0.11 & $0.32\times0.28$, $-45.8^\circ$ \\
W1248$-$2154 $^b$ & CO(4-3) & 28-01-2018 & J1337$-$1257 & J1245$-$1616 & 0.12 & $0.50\times0.48$, 46.9$^\circ$ \\ [2pt]
W1603+2745 & CO(3-2) & 24-01-2018 & J1550+0527 & J1619+2247 & 0.16 & $1.09\times0.74$, $-19.7^\circ$ \\
           &        & 30-09-2018 & J1550+0527 & J1619+2247 &  \\ [2pt]
W1814+3412 & CO(4-3) & 10-01-2018 & J1751+0939 & J1753+2848 & 0.32 & $0.61\times0.46$, 22.6$^\circ$ \\
W2054+0207 & CO(4-3) & 20-01-2018 & J2134$-$0153 & J2101+0341 & 0.23 & $0.74\times0.52$, 50.8$^\circ$ \\ 
W2201+0226 & CO(3-2) & 01-01-2018 & J2148+0657 & J2156-0037 & 0.39 & $0.62\times0.51$, $-46.7^\circ$ \\
W2210$-$3507 & CO(3-2) & 01-01-2018 & J2258$-$2758 & J2151$-$3027 & 0.35 & $0.68\times0.60$, $-58.4^\circ$ \\
W2238+2653 & CO(4-3) & 01-10-2018 & J2253+1608 & J2236+2828 & 0.35 & $0.72\times0.51$, $-4.3^\circ$ \\
W2305$-$0039 $^b$ & CO(4-3) & 10-12-2017 & J0006$-$0623 & J2301$-$0158 & 0.12 & $0.39\times0.32$, 60.5$^\circ$ \\
\enddata
\tablecomments{$^a$ Sensitivity in a 100\,km\,s$^{-1}$ velocity bin.
$^b$ From project 2017.1.00358.S; we note that W1248–2154 was observed in both projects, however we use these data as they are deeper.}
\end{deluxetable*}

\section{ALMA observations and data analysis} \label{sec:alma}
Observations were carried out with ALMA using the Band-3 and Band-4 receivers during Cycle 5. 
Most of our sample sources were observed in our project 2017.1.00441.S (PI:
L.~Fan). A few sources were allocated to a different project and available in
the ALMA archive (2017.1.00358.S), and we also include these observations.
\footnote{We note that 2017.1.00358.S has more sources; however, as these were not part of our original sample, we do not include them in this analysis.} 
Table \ref{tab:obssum} summarizes the details of the observations, including a
list of the calibrators.  We note that while the observations in 2017.1.00441.S
used a spectral setup for the spectral window (spw) of the sideband that was
predicted to include the emission line (and continuum mode for the spectral
windows of the other sideband), project 2017.1.00358.S used a continuum
setup for all spectral windows.

Reduction, calibration, and imaging were done using {\sc casa} (Common
Astronomy Software Application\footnote{\url{https://casa.nrao.edu}};
\citealt{Mcmullin07}).  
The pipeline-reduced data delivered from the observatory was of sufficient
quality such that no additional flagging and further calibration were necessary.  
The pipeline includes the steps required for a standard reduction and
calibration, such as flagging, bandpass calibration, as well
as flux and gain calibration.  
A conservative estimate of the uncertainty of the absolute flux calibration is $10\%$. 

The data was imaged both as continuum and spectral cube using natural
weighting. The {\sc casa} task {\tt uvcontsub} was
used to subtract the continuum from {\it uv} data for sources for which the continuum was detected.  
A continuum image was produced combining all spectral windows,
while a spectral cube was constructed for the spectral windows tuned
to the redshifted CO line.  
The rms sensitivity and the synthesized beam size achieved by imaging with a natural weighting scheme are given in Table~\ref{tab:obssum}.

\section{Multiwavelength Data and Spectral Energy Distribution fitting}\label{section3}
\subsection{UV to Millimeter Spectral Energy Distribution Data}
To decompose the host galaxy emission from the central AGN and estimate their physical properties, such as stellar mass $(M_\star)$ and SFR, we constructed UV to millimeter
SEDs for all objects in our sample. Various photometry catalogs were retrieved. 13 Hot DOGs in our sample have optical to near-infrared (NIR) broadband photometric data from different surveys, including the first public data release of the Dark Energy Survey \citep[DES DR1;][]{desdr1} \footnote{\url{https://des.ncsa.illinois.edu/releases/dr1/}} in the g, r, i, z, and Y bands, 
the seventh public data release of the Dark Energy Camera Legacy Survey \citep[DECaLS DR7;][]{Dey2019} \footnote{\url{https://www.legacysurvey.org/dr7/}} in the g, r, and z bands, the third Data Release of the Beijing-Arizona Sky Survey \citep[BASS DR3;][]{Zou2019} \footnote{\url{http://explore.china-vo.org/data/bassdr3coadd/}} in the g, r, and z bands, the fourth Data Release of Kilo-Degree Survey \citep[KiDs DR4;][]{Kuijken2019} \footnote{\url{http://kids.strw.leidenuniv.nl/DR4/}} together with the Visible and Infrared Survey Telescope for Astronomy (VISTA) Kilo-degree Infrared Galaxy (VIKING) Survey \citep{Edge2013} in the u, g, r, i, Z, Y, J, H, and $\rm K_s$ bands, Two Micron All Sky Survey (2MASS) photometry from NED \footnote{\url{https://ned.ipac.caltech.edu/}} in the H, and $\rm K_s$ bands and the Sloan Digital Sky Survey (SDSS) \footnote{\url{https://www.sdss.org/dr15/}}. The 3 remaining Hot DOGs, namely W0248+2705, W0615-5716, and W1248-2154, currently lack optical-NIR data. The WISE W1 and W2 
flux densities were obtained through aperture photometry on the WISE images \footnote{\url{https://unwise.me/}} \citep[from the unWISE catalog;][]{Lang2014,Meisner2017}, 
and the errors were estimated based on the inverse variance images. The optical-NIR photometry catalog of our sample is shown in Table \ref{tab:SED}. The WISE 
W3 and W4 photometry data were obtained from ALLWISE Data Release \citep{Cutri2013}. For the far-infrared(FIR)tomillimeter photometry, because of the sample selection in \citet{Fan2016b}, all Hot DOGs have \textit{Hershcel} Photoconductor Array Camera and Spectrometer \citep[PACS;][]{Poglitsch2010} observations at 70 and 
160 $\mu$m and Spectral and Photometric Imaging REceiver \citep[SPIRE;][]{Griffin2010} observations at 250, 350, and 500 $\mu$m \citep{Pilbratt2010}. Part of our sample 
have James Clerk Maxwell Telescope (JCMT) SCUBA-2 observations \citep{Jones2014}, as well as CSO SHARC-II observations 
at 850$\mu$m, CSO Bolocam observations at 1.1 mm \citep{Wu2012}, and Submillimeter Array (SMA) observations at 1.3 
mm \citep{Wu2014}. The IR broadband photometry at wavelengths ranging from 12 $\mu$m to the millimeter band were directly collected 
from their parent samples as reported in \citet{Fan2016b}. The ALMA continuum observations at rest-frame 3 mm have been 
included to constrain the cold dust component more accurately.

\begin{deluxetable*}{lccccccccccc}
\tablecaption{Optical-NIR photometry. \label{tab:SED}}
\tabletypesize{\scriptsize}
\setlength{\tabcolsep}{2pt}
\tablehead
{\\
\colhead{Source}  &  \colhead{u}    &   \colhead{g}    &   \colhead{r}    &   \colhead{i} & \colhead{z} & \colhead{Y} & \colhead{J} & \colhead{H} & \colhead{Ks} & \colhead{W1} & \colhead{W2}\\
    &  \colhead{ $\mu$Jy}    &  \colhead{ $\mu$Jy}    &  \colhead{ $\mu$Jy}    &  \colhead{ $\mu$Jy}    &  \colhead{ $\mu$Jy}    &  \colhead{ $\mu$Jy}    &  \colhead{ $\mu$Jy}    &  \colhead{ $\mu$Jy}    &  \colhead{ $\mu$Jy}    &  \colhead{ $\mu$Jy}    &  \colhead{ $\mu$Jy}\\}
\startdata
W0134-2922$^a$  & $1.27\pm0.34$ & $3.54\pm0.11$ & $4.53\pm0.12$ & $6.13\pm0.84$ & $6.23\pm0.38$ & $8.84\pm0.9$ & $8.88\pm0.81$ & $10.77\pm1.97$ & $16.38\pm1.85$ & $23.04\pm4.7$ & $96.21\pm9.68$\\  
W0149+2350$^b$  & .. & $0.45\pm0.15$ & $1.21\pm0.3$ & .. & $2.57\pm0.44$ & .. & .. & .. & .. & $14.47\pm2.09$ & $25.88\pm4.78$\\
W0220+0137$^{ce}$  & $1.49\pm0.85$ & $7.06\pm0.38$ & $6.63\pm0.53$ & $5.42\pm0.66$ & $8.16\pm2.06$ & .. & .. & .. & .. & $18.53\pm2.06$ & $31.55\pm4.84$\\ 
W0248+2705  & .. & .. & .. & .. & .. & .. & .. & .. & .. & $10.7\pm4.99$ & $30.48\pm10.77$\\
W0410-0913$^b$  & .. & $0.32\pm0.26$ & $2.57\pm0.42$ & .. & $2.55\pm0.52$ & .. & .. & .. & .. & $65.88\pm2.2$ & $66.21\pm4.8$\\
W0533-3401$^c$  & .. & $4.86\pm0.21$ & $7.12\pm0.27$ & $9.62\pm0.52$ & $13.14\pm1.11$ & $11.64\pm3.21$ & .. & .. & .. & $35.07\pm1.65$ & $72.61\pm3.63$\\
W0615-5716  & .. & .. & .. & .. & .. & .. & .. & .. & .. & $35.52\pm2.91$ & $43.33\pm5.24$\\
W1248-2154  & .. & .. & .. & .. & .. & .. & .. & .. & .. & $44.42\pm4.59$ & $35.48\pm9.35$\\
W1603+2745$^b$  & .. & $0.46\pm0.13$ & $0.82\pm0.19$ & .. & $1.95\pm0.36$ & .. & .. & .. & .. & $7.42\pm1.67$ & $31.08\pm3.86$\\
W1814+3412$^d$  & .. & $1.33\pm0.2$ & $6.51\pm0.57$ & .. & $7.42\pm0.71$ & .. & .. & .. & .. & $8.63\pm3.59$ & $15.16\pm7.11$\\
W2054+0207$^b$  & .. & $0.83\pm0.12$ & $1.5\pm0.15$ & .. & $5.48\pm0.4$ & .. & .. & .. & .. & $14.83\pm2.03$ & $115.43\pm4.64$\\
W2201+0226$^b$  & .. & $0.85\pm0.11$ & $1.05\pm0.22$ & .. & $2.34\pm0.68$ & .. & .. & .. & .. & $18.2\pm2.17$ & $83.29\pm5.01$\\
W2210-3507$^a$  & $0.84\pm0.25$ & $1.17\pm0.09$ & $1.68\pm0.1$ & $1.69\pm0.5$ & $3.05\pm0.47$ & $2.15\pm0.88$ & $6.88\pm0.75$ & $8.72\pm1.7$ & $13.61\pm2.23$ & $27.49\pm4.96$ & $34.8\pm11.38$\\
W2238+2653$^{be}$  & $1.1\pm0.76$ & $1.51\pm0.32$ & $2.85\pm0.54$ & $5.06\pm0.84$ & $5.3\pm2.78$ & .. & .. & .. & .. & $27.45\pm1.9$ & $59.72\pm4.32$\\
W2246-0526$^f$  & .. & .. & .. & .. & .. & .. & .. & $5.2\pm0.2$ & $8.8\pm2.8$ & $31.0\pm7.0$ & $34.0\pm7.0$\\
W2305-0039$^{ce}$  & $0.59\pm0.47$ & .. & $1.98\pm0.5$ & $3.97\pm0.72$ & $6.25\pm2.47$ & .. & .. & .. & .. & $22.83\pm2.27$ & $44.56\pm5.25$\\
\enddata
\tablecomments{$^a$ (u, g, r, i, z, Y, J, H, Ks) bands photometry from KiDS DR4 and VIKING catalog. $^b$ (g, r, z) bands photometry from DECaLS DR7 catalog. $^c$ (g, r, i, z, Y) bands photometry from DES DR1 catalog. $^d$ (g, r, z) bands photometry from BASS DR3 catalog. $^e$ (u, i) bands photometry from SDSS. $^f$ (H, Ks) bands from 2MASS photometry.}
\end{deluxetable*}

\begin{deluxetable*}{lccccccccc}
\tablecaption{CO and Millimeter Continuum Measurements Based on ALMA Observations. \label{tab:results}}
\tabletypesize{\scriptsize}
\tablehead{
\\
\colhead{Source} & \colhead{R.A.} & \colhead{Dec} & \colhead{$z_{\rm CO}$} & \colhead{$S_{\rm peak}$} & \colhead{FWHM} & \colhead{$I_{\rm CO}$} & \colhead{$\nu_{\rm cen}$} & \colhead{$S_{\rm cont}$} & \colhead{size}\\
\colhead{Name} & \colhead{(J2000)}  & \colhead{(J2000)} & \colhead{} & \colhead{(mJy/beam)} & \colhead{(km\,s$^{-1}$)}  &  \colhead{(Jy\,km\,s$^{-1}$)} & \colhead{GHz}  & \colhead{($\mu$Jy)} & \colhead{($''\times''$)}\\
\colhead{(1)} & \colhead{(2)} & \colhead{(3)} & \colhead{(4)} & \colhead{(5)} & \colhead{(6)} & \colhead{(7)} & \colhead{(8)} & \colhead{(9)} & \colhead{(10)}
}

\startdata
W0126$-$0529$^a$$^b$ & .. & .. & .. & $< 1.35$ & 300 & $<0.43$ & 93.64 & $462\pm71$ & $0.41\times0.33$, 30$^\circ$ \\
W0134$-$2922 & 01:34:35.70 & $-$29:22:45.54 & $3.0572\pm0.0004$ & $0.85\pm0.09$ & $612\pm 74$ & $0.55\pm0.09$ & 106.96 & $35\pm10$ & unresolved \\
W0248+2705TT & 02:48:58.71 & +27:05:30.08 & $2.1825\pm0.0007$ & $1.26\pm0.29$ & $551\pm150$ & $0.74\pm0.26$ \\
W0533$-$3401 & 05:33:58.41 & $-$34:01:34.50 & $2.9024\pm0.0003$ & $4.83\pm0.42$ & $555\pm 55$ & $2.86\pm0.38$  & 94.36 & $169\pm48$ & unresolved \\
W0615$-$5716 & 06:15:11.10 & $-$57:16:14.81 & $3.3463\pm0.0007$ & $0.54\pm0.08$ & $723\pm115$ & $0.42\pm0.09$ &  99.70 & $<60$  & ..   \\
W1248$-$2154 & 12:48:15.20 & $-$21:54:20.12 & $3.3233\pm0.0007$ & $0.31\pm0.05$ & $610\pm113$ & $0.20\pm0.05$ & 99.94 & $147\pm28$ & unresolved \\ 
W1603+2745 & 16:03:57.36 & +27:45:52.95 & $2.6540\pm0.0005$ & $1.77\pm0.15$ & $1081\pm103$ & $2.04\pm0.26$ & 100.88 & $94\pm27$ & extended \\
W1814+3412 TT & 18:14:17.27 & +34:12:24.45 & $2.4568\pm0.0004$ & $0.76\pm0.19$ & $254\pm 74$ & $0.21\pm0.08$ & 139.12 & $<125$ & .. \\
%W1814+3412 & .. & .. & .. & $<0.96$ & 300 & $<0.31$ & $<0.6\times10^{10}$ \\
W2054+0207 & 20:54:25.68 & +02:07:11.56 & $2.5323\pm0.0002$ & $4.87\pm0.31$ & $457\pm 34$ & $2.37\pm0.23$ & 136.61 & $212\pm52$ & $0.55\times0.41$, 165$^\circ$ \\
W2201+0226 & 22:01:23.38 & +02:26:21.87 & $2.8752\pm0.0003$ & $15.0\pm0.87$ & $691\pm 46$ & $11.0\pm0.97$ & 94.87 & $119\pm36$ & unresolved \\
W2210$-$3507$^a$ & .. & .. & .. & $<1.1$ & 300 & $<0.34$ & 96.35 & $<130$ & .. \\
W2238+2653 & 22:38:10.19 & +26:53:20.01 & $2.3987\pm0.0002$ & $13.0\pm0.52$ & $698\pm 32$ & $9.65\pm0.58$ & 141.02 & $484\pm82$ & $0.77\times0.32$, 170$^\circ$ \\
W2305$-$0039 & 23:05:25.86 &  $-$00:39:25.35 & $3.1107\pm0.0002$ & $5.79\pm0.23$ & $570\pm 26$ & $3.51\pm0.21$ & 107.246 &  $220\pm22$ & $0.22\times0.16$, 177$^\circ$ \\
\enddata
\tablecomments{(1): Source names. $^a$ 3$\sigma$ upper limits for CO line intensity are calculated assuming a linewidth of 300\,km\,s$^{-1}$ and $5\sigma$ upper limits for the continuum are calculated assuming the source is compact. $^b$ Continuum position 01:26:11.95 $-$05:29:09.31. TT: tentative line detection. (2) and (3): Positions derived from the moment-0 maps. (4): Redshift based on CO (3$-$2) emission line. (5), (6), and (7): Peak flux density, line width, and line intensity based on Gaussian line profile. (8), (9), and (10): The frequency, flux density, and size of the continuum detections based on a 2D Gaussian profile fitting. }
\end{deluxetable*}

\subsection{SED Analysis} \label{sec:3.2}
The UV to millimeter SED fitting in our study was conducted with the latest version of BayeSED \citep{Han2012a,HanY2014a,HanY2019a} \footnote{\url{https://bitbucket.org/hanyk/bayesed/}}, namely BayeSED V3.0. This new version has improved the accuracy and speed of the stellar population synthesis algorithm and has been tested with mock galaxies to show good quality and speed for parameter estimation of galaxies \citep{HanY2023a}. For SED models given as a SED library, principal component analysis is employed to reduce memory usage. Then, interpolation between the SED models is conducted with artificial neural networks or K-nearest neighbors to allow a fast and continuous sampling of high-dimensional parameter space. Finally, the MultiNest algorithm is employed to sample the parameter space and calculate the posterior probability distribution of the parameters.

The SED of each galaxy in our sample was decomposed into three components: stellar, cold dust, and central AGN. The stellar emission was modeled by using the \citet{bc03} 
simple stellar population library with a \citet{chabrier2003} initial mass 
function (IMF), and an exponential declining star formation history (SFH). The \citet{Calzetti2000a} 
dust attenuation law was used, and the energy of stellar emission absorbed by cold dust was assumed to be totally reemitted in the IR band. This assumption can break the degeneracy in the UV and optical bands by the complement of IR photometry \citep{dacunha2008,Buat2014}, and stellar population properties can be more robustly constrained. The cold dust emission was modeled 
conventionally as a graybody, which was defined as $S_{\lambda}\propto(1-e^{-(\frac{\lambda_0}
{\lambda})^{\beta}}) B_\lambda(T_{\rm dust})$, where $\lambda_0$=125$\mu$m, and the dust temperature $T_{\rm dust}$ and the  
emissivity index $\beta$ are two free parameters in the SED fitting. The CLUMPY torus model \citep{Nenkova2008a,Nenkova2008b} \footnote{\url{http://www.pa.uky.edu/clumpy/models/clumpy_models_201410_ tvavg.hdf5/}} has significantly advanced the modeling of IR emission in various AGN samples \citep{Ramos2017}, and has been utilized to model the UV-IR emission of the central AGN of our samples. Six parameters are employed in the CLUMPY model to describe the geometry and physical properties of the CLUMPY torus. The SED analysis method employed in this study is consistent with that used in a previous work by \citet{Fan2019}, who presented a comprehensive analysis of W0533-3401. However, we have restricted the $T_{\rm dust}$ parameter to a range of 30-50K to address AGN contamination in the FIR, as discussed in Section \ref{sec:5.4}. The 12 free parameters used in the fitting process are listed in Table 3 of \citet{Fan2019}. For the three Hot DOGs without optical-NIR photometry, we excluded the stellar population component to prevent overfitting. Thus, the stellar population parameters, such as $M_\star$, have not been estimated for these three galaxies. The upper flux
density limit was taken into account by setting the corresponding flux density error to a negative value according to the convention in BayeSED.

\section{Results}\label{section4}

\subsection{CO Emission Lines and Continuum}

\begin{figure*}
\includegraphics[width=0.2465\textwidth]{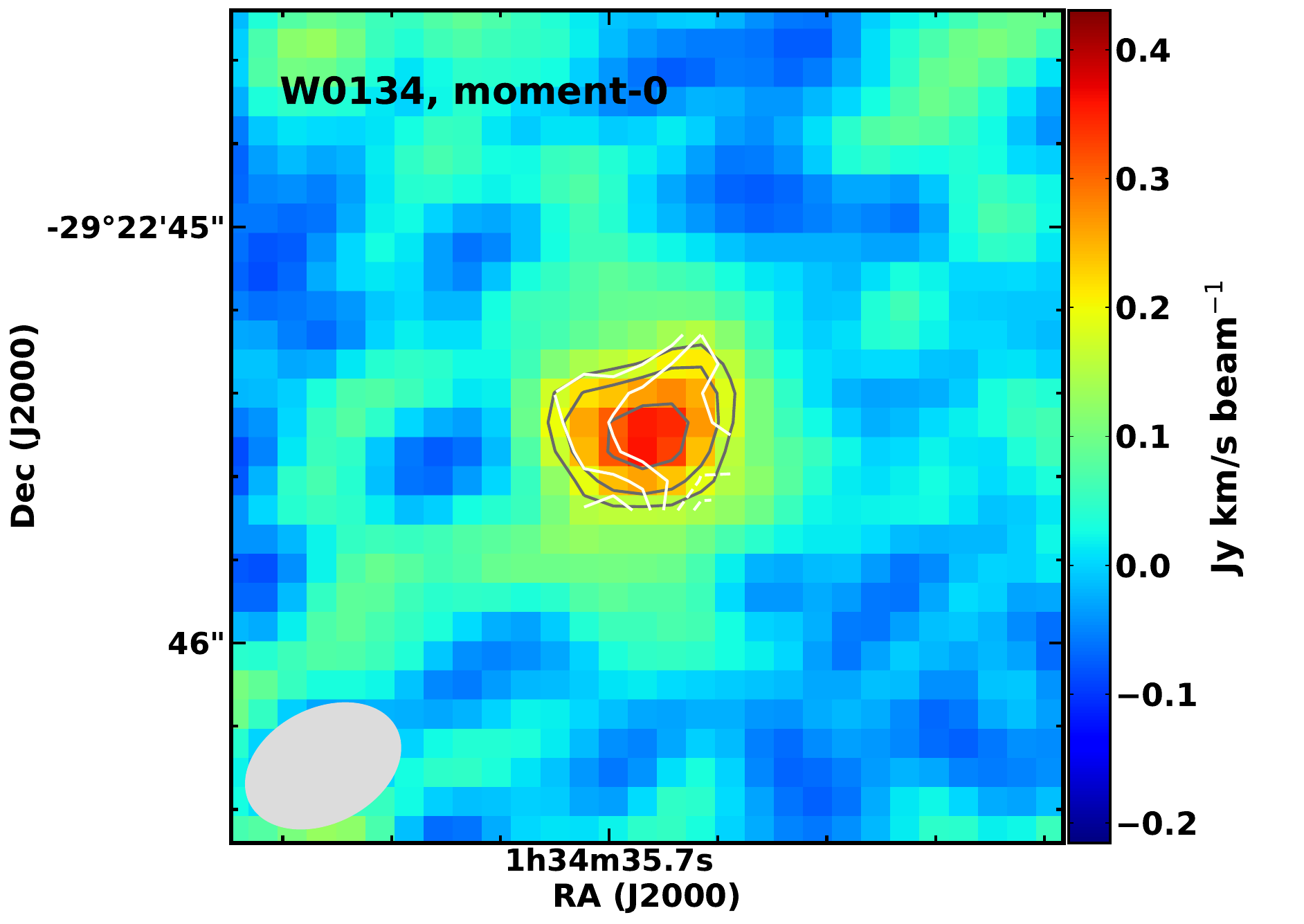}
\includegraphics[width=0.2465\textwidth]{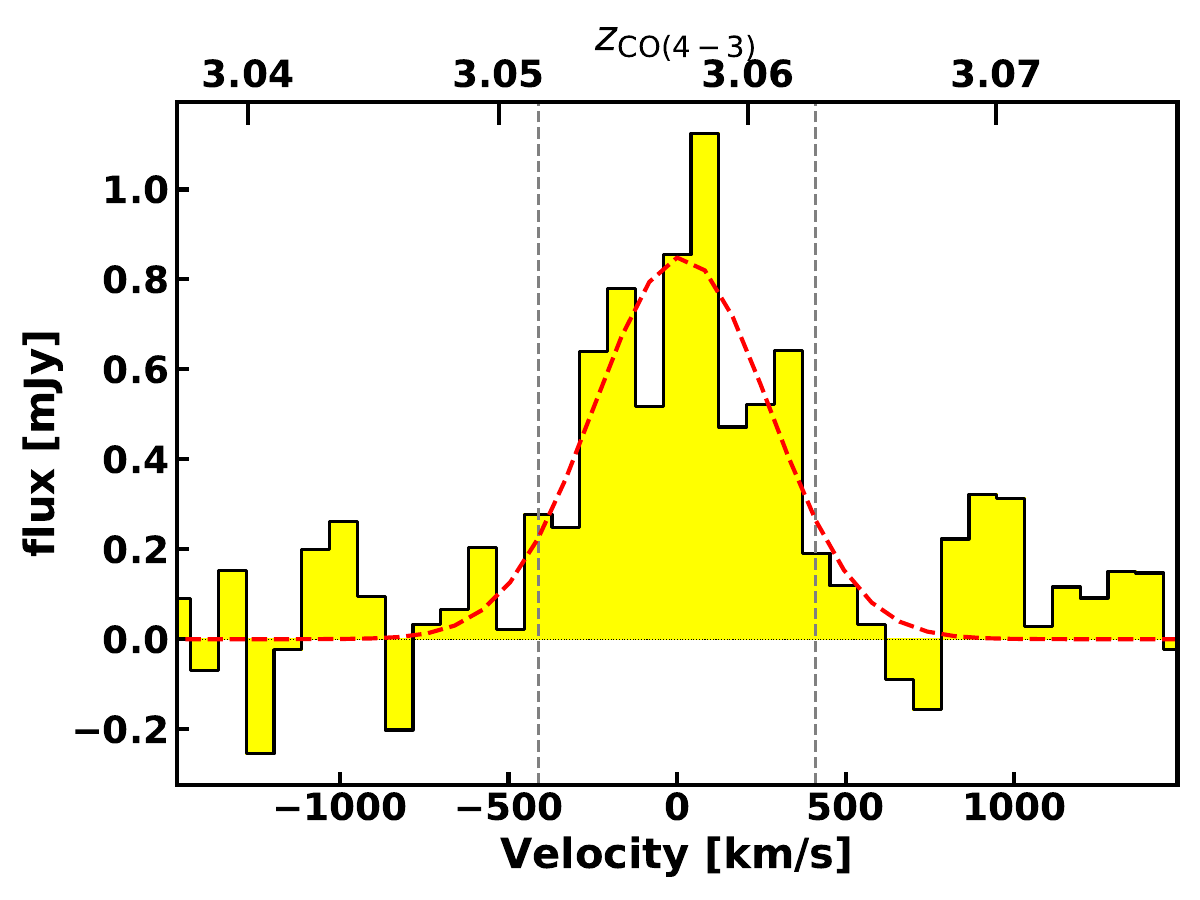}
\includegraphics[width=0.2465\textwidth]{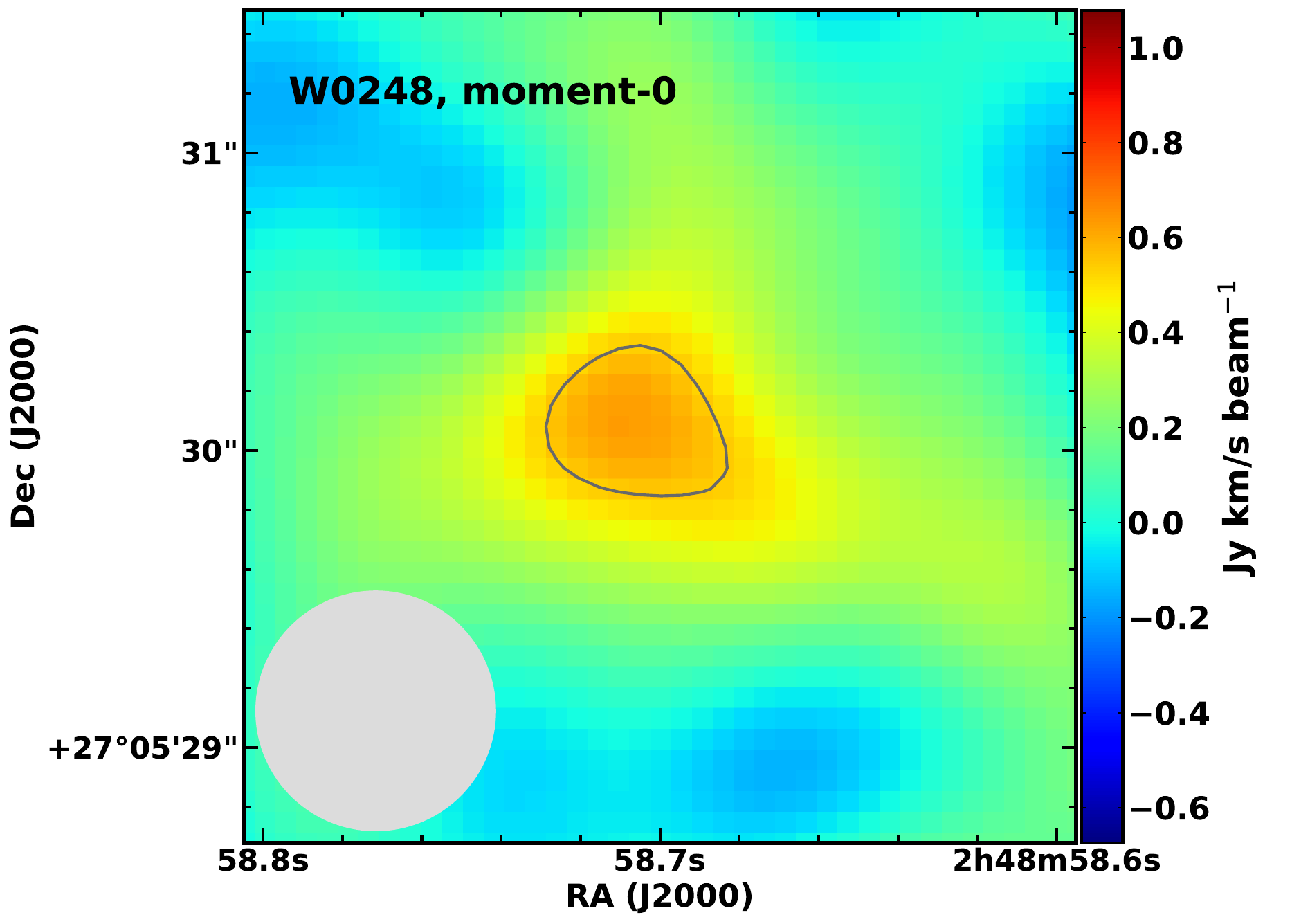}
\includegraphics[width=0.2465\textwidth]{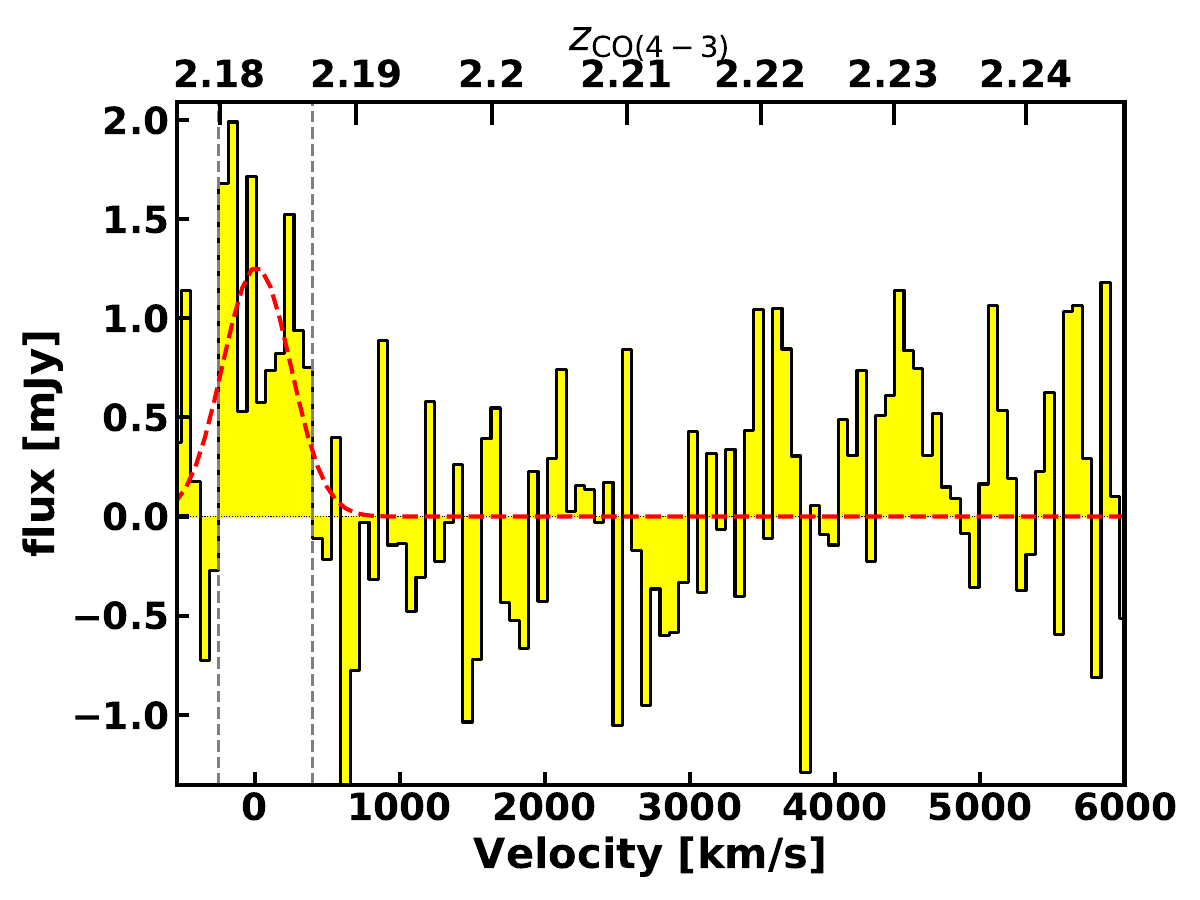}
\includegraphics[width=0.2465\textwidth]{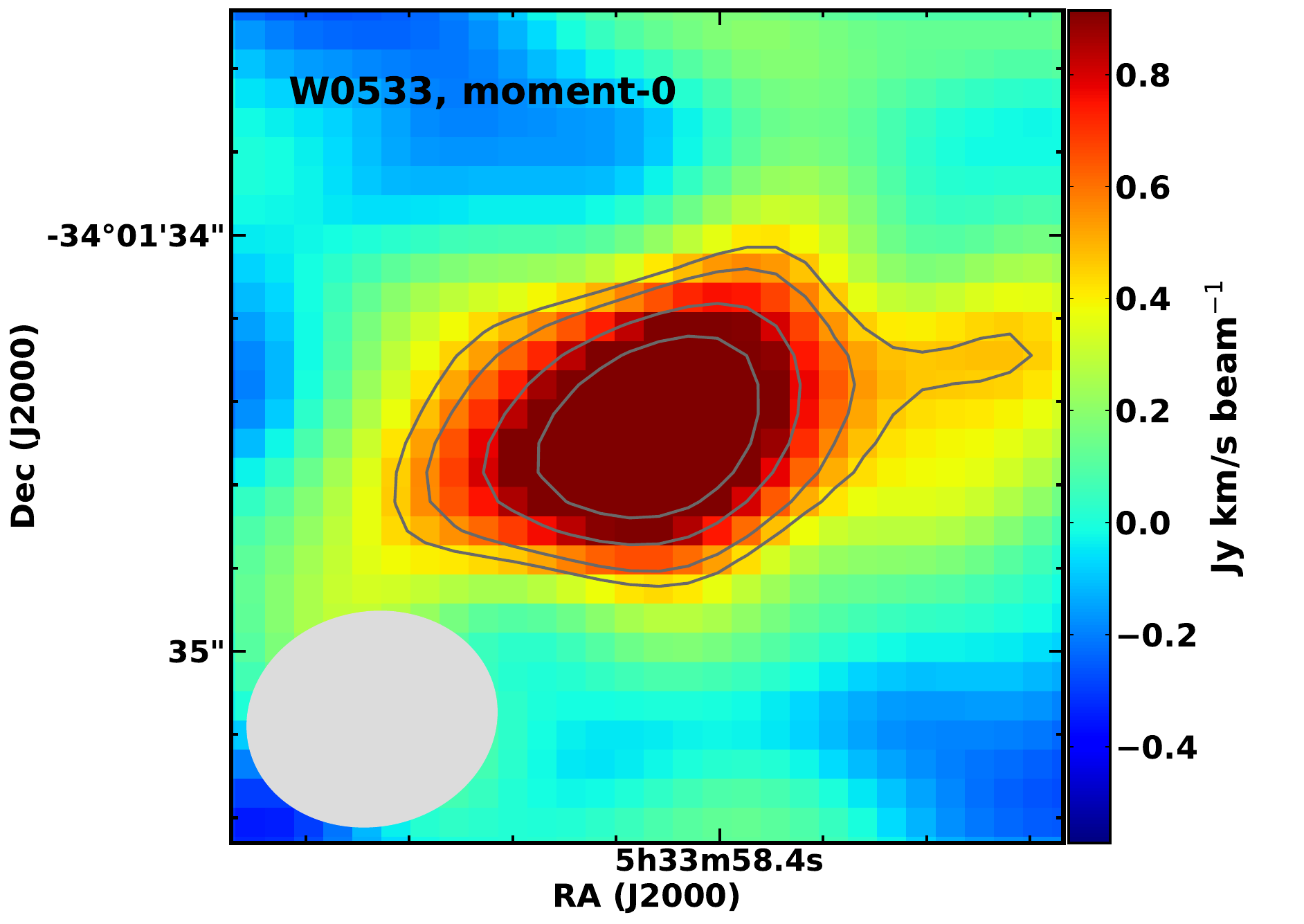}
\includegraphics[width=0.2465\textwidth]{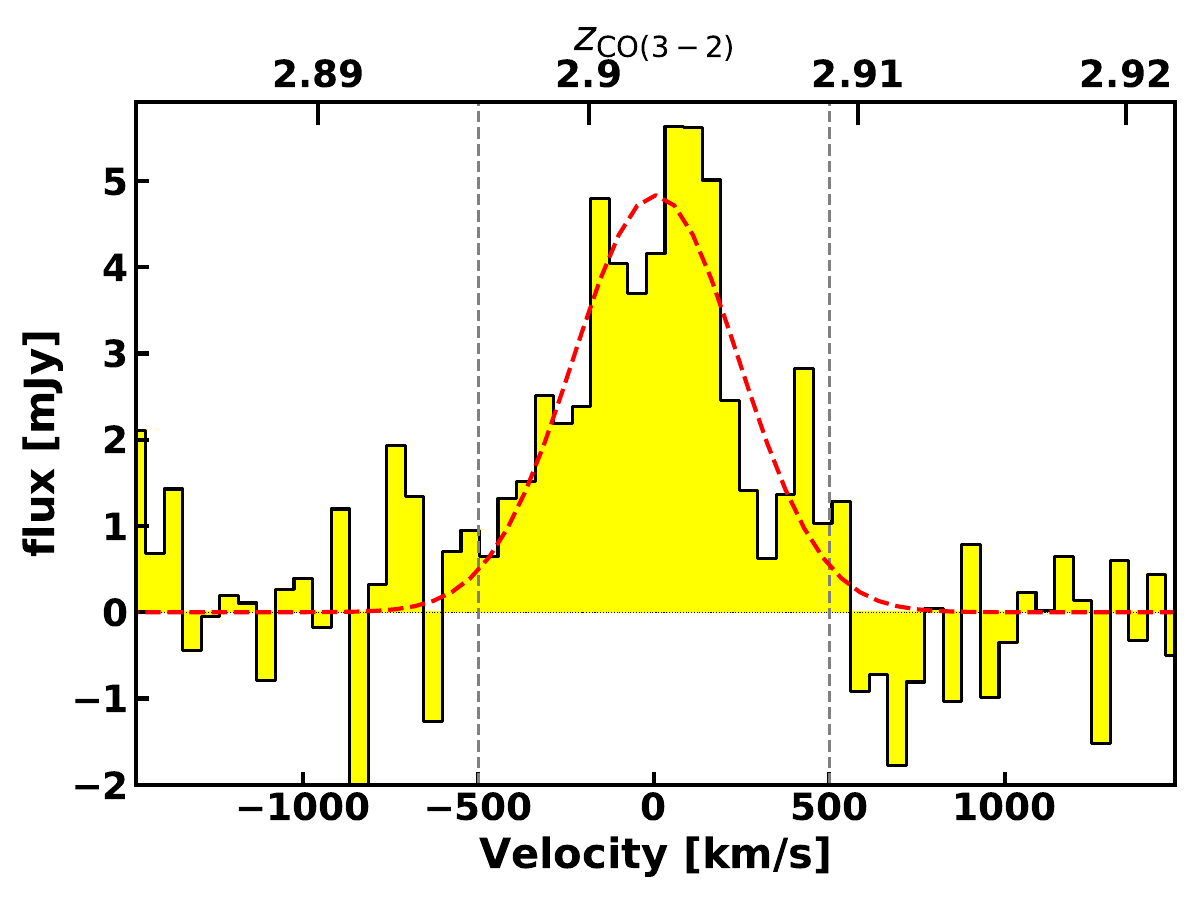}
\includegraphics[width=0.2465\textwidth]{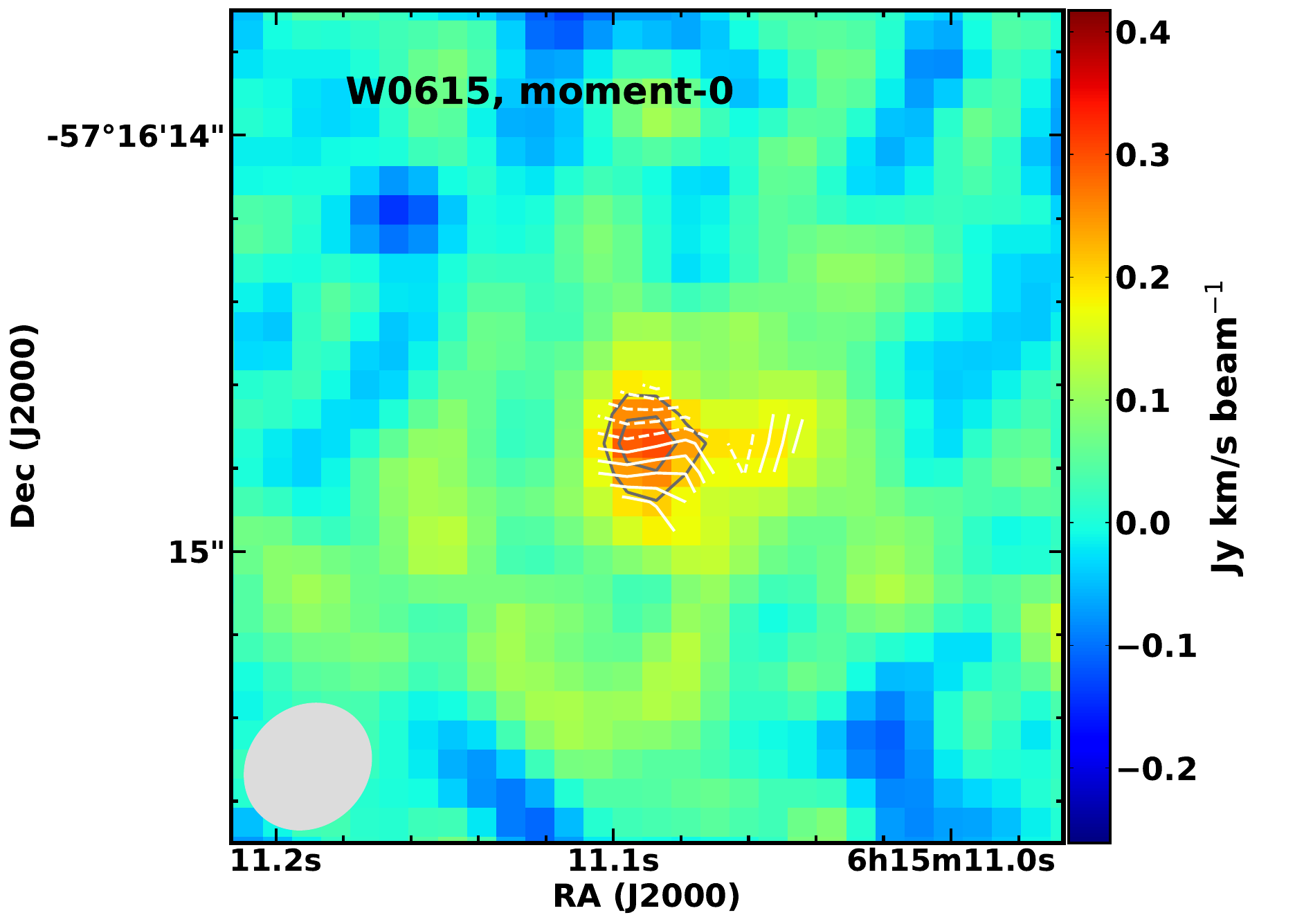}
\includegraphics[width=0.2465\textwidth]{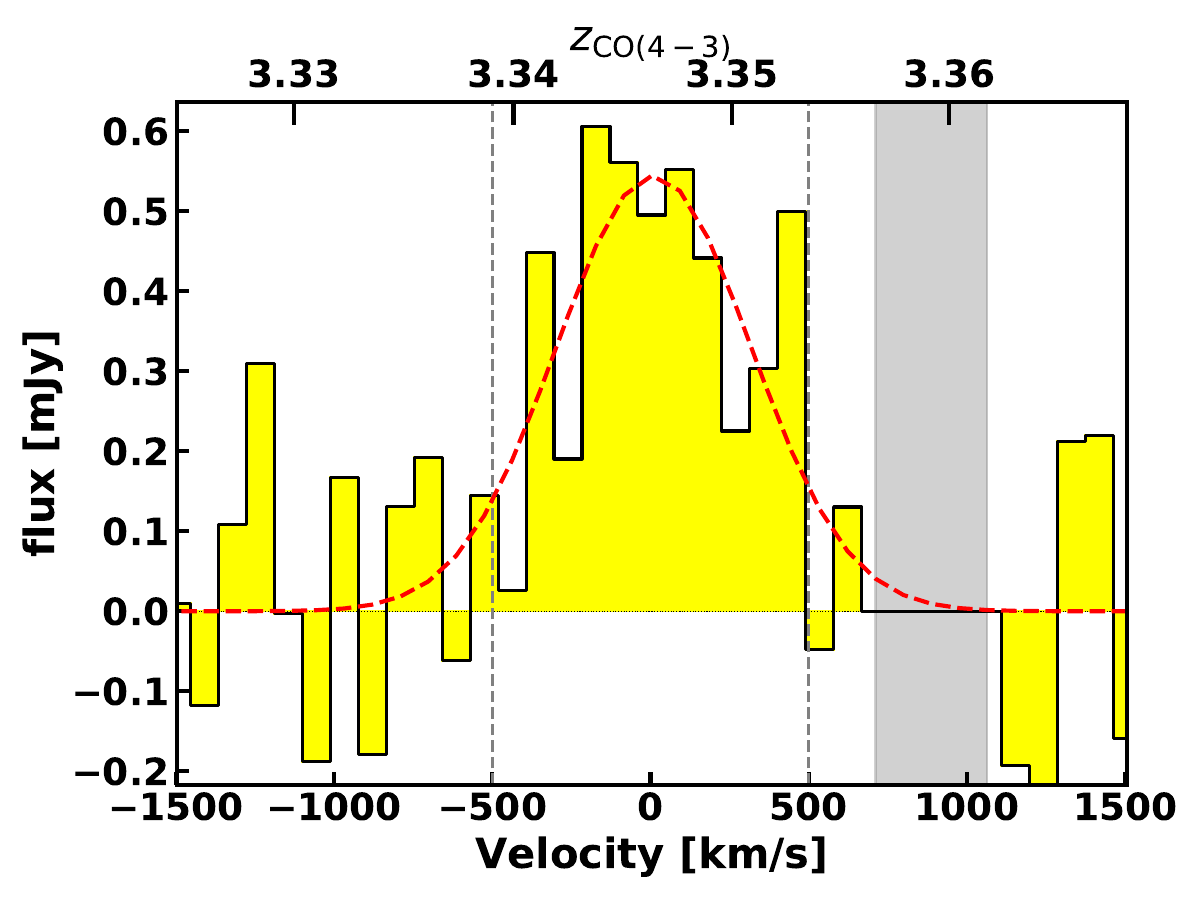}
\includegraphics[width=0.2465\textwidth]{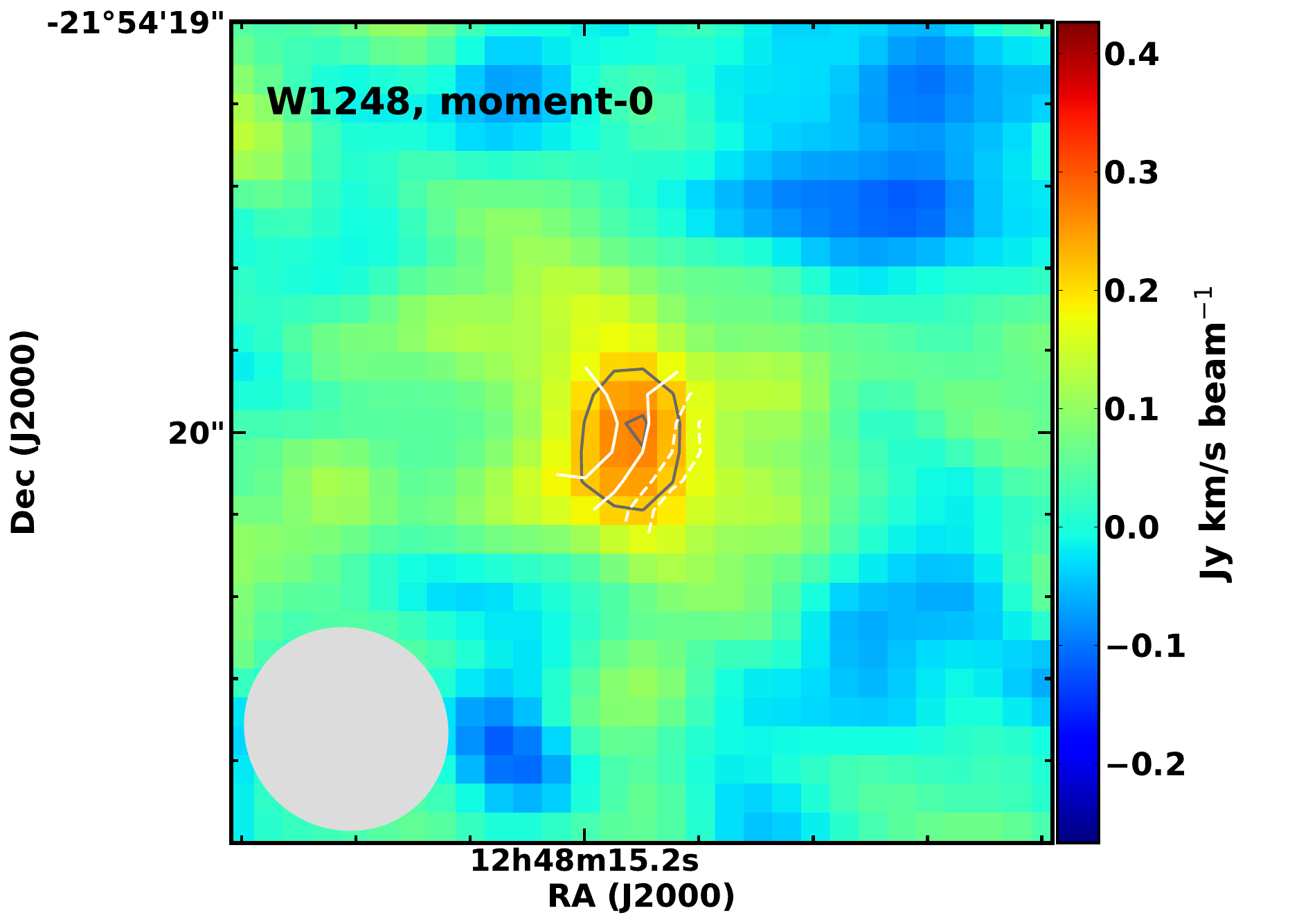}
\includegraphics[width=0.2465\textwidth]{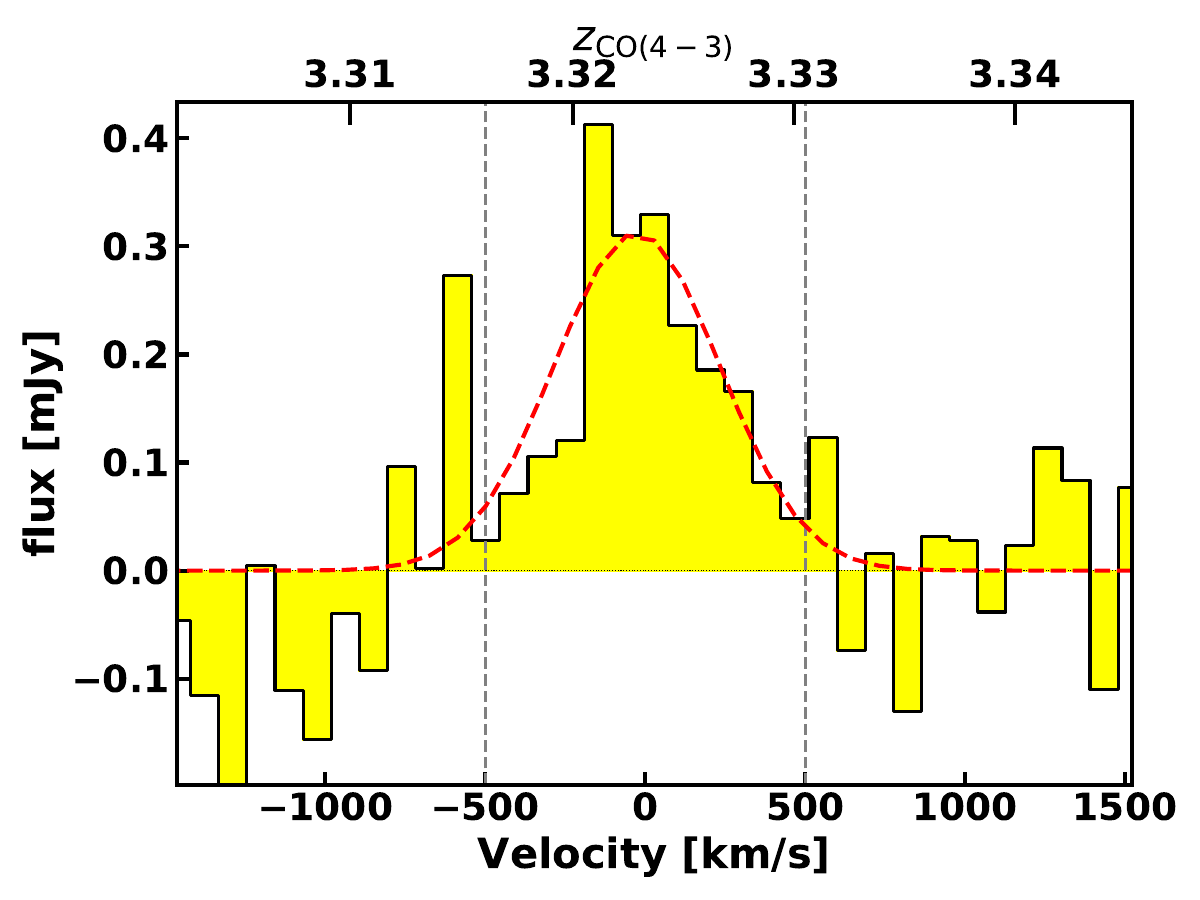}
\includegraphics[width=0.2465\textwidth]{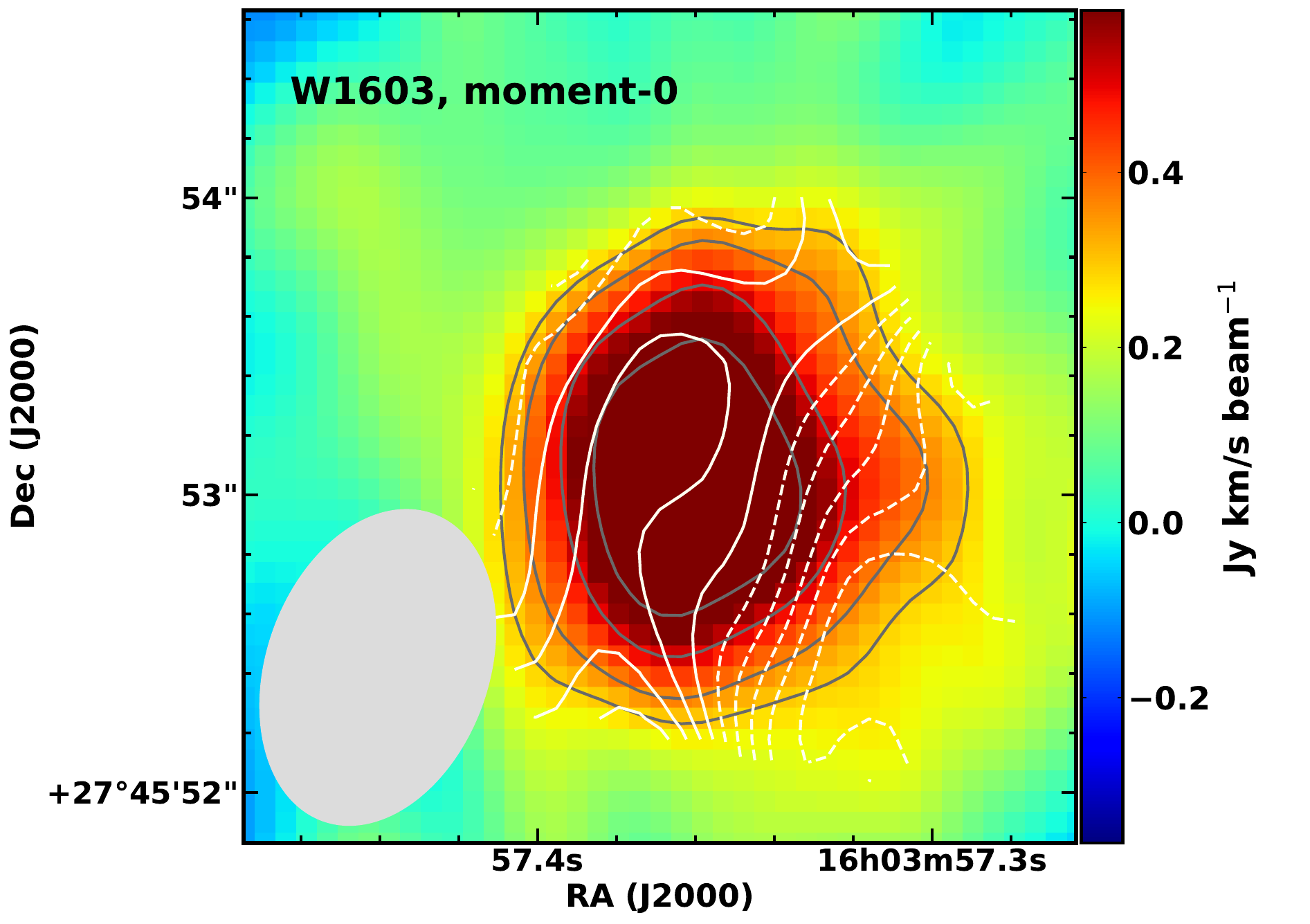}
\includegraphics[width=0.2465\textwidth]{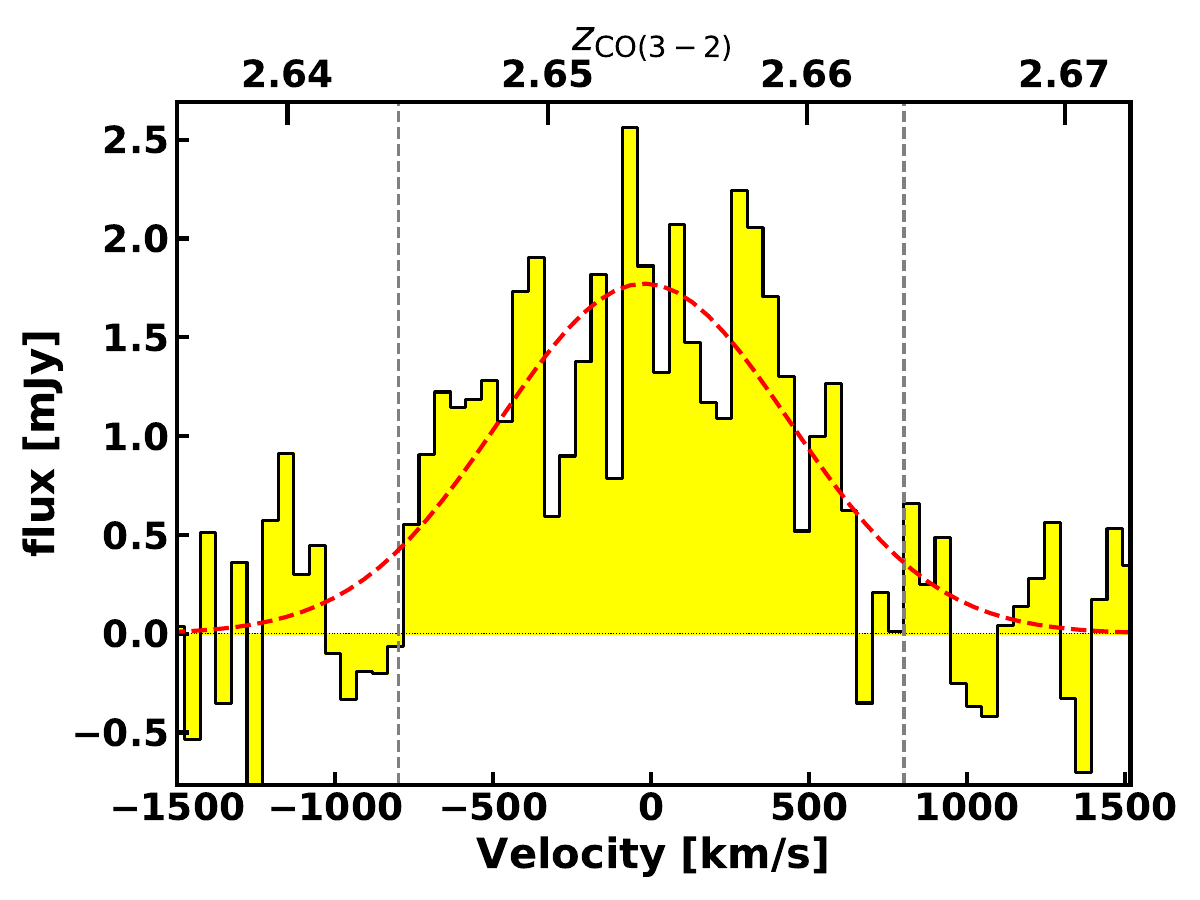}
\includegraphics[width=0.2465\textwidth]{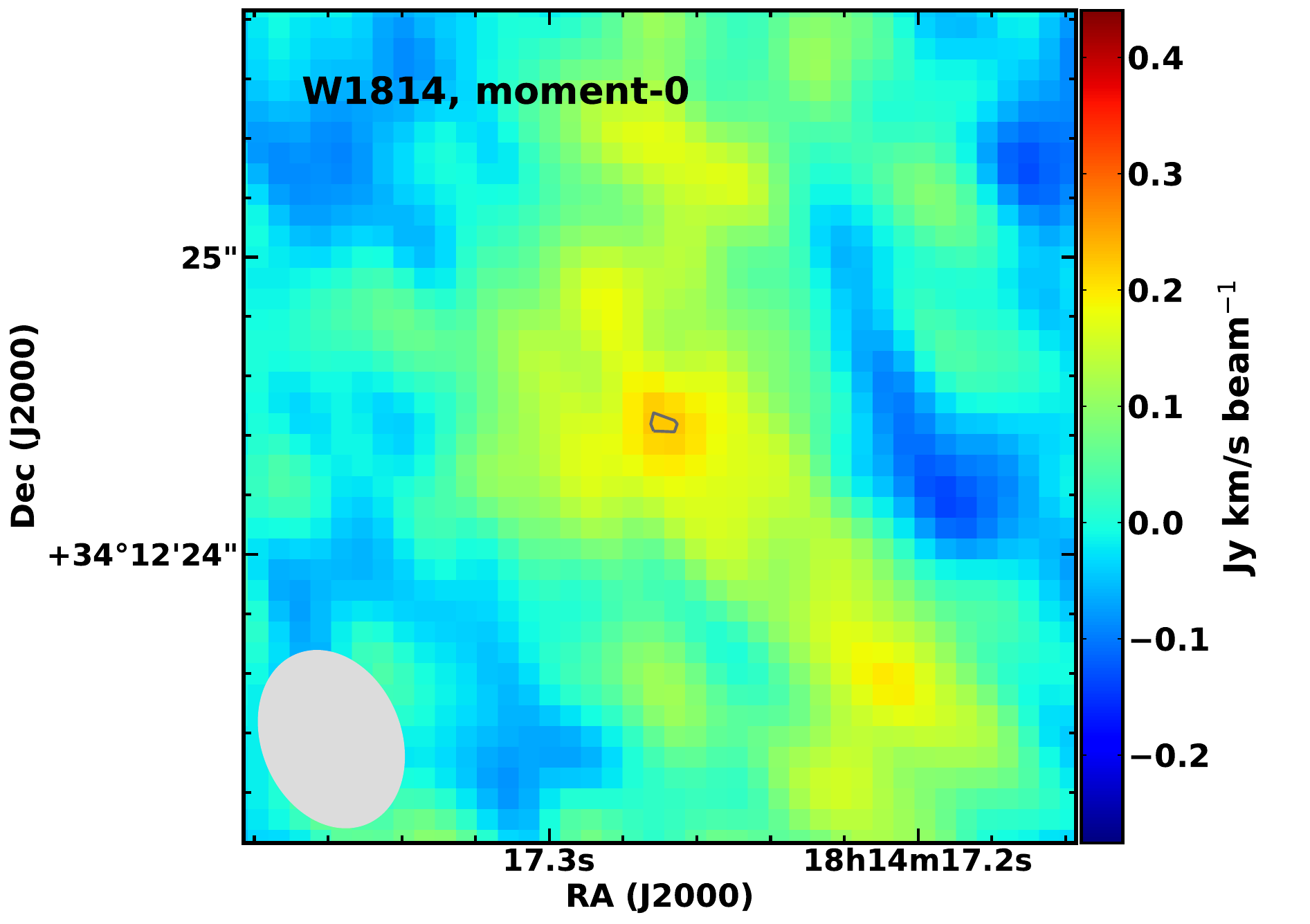}
\includegraphics[width=0.2465\textwidth]{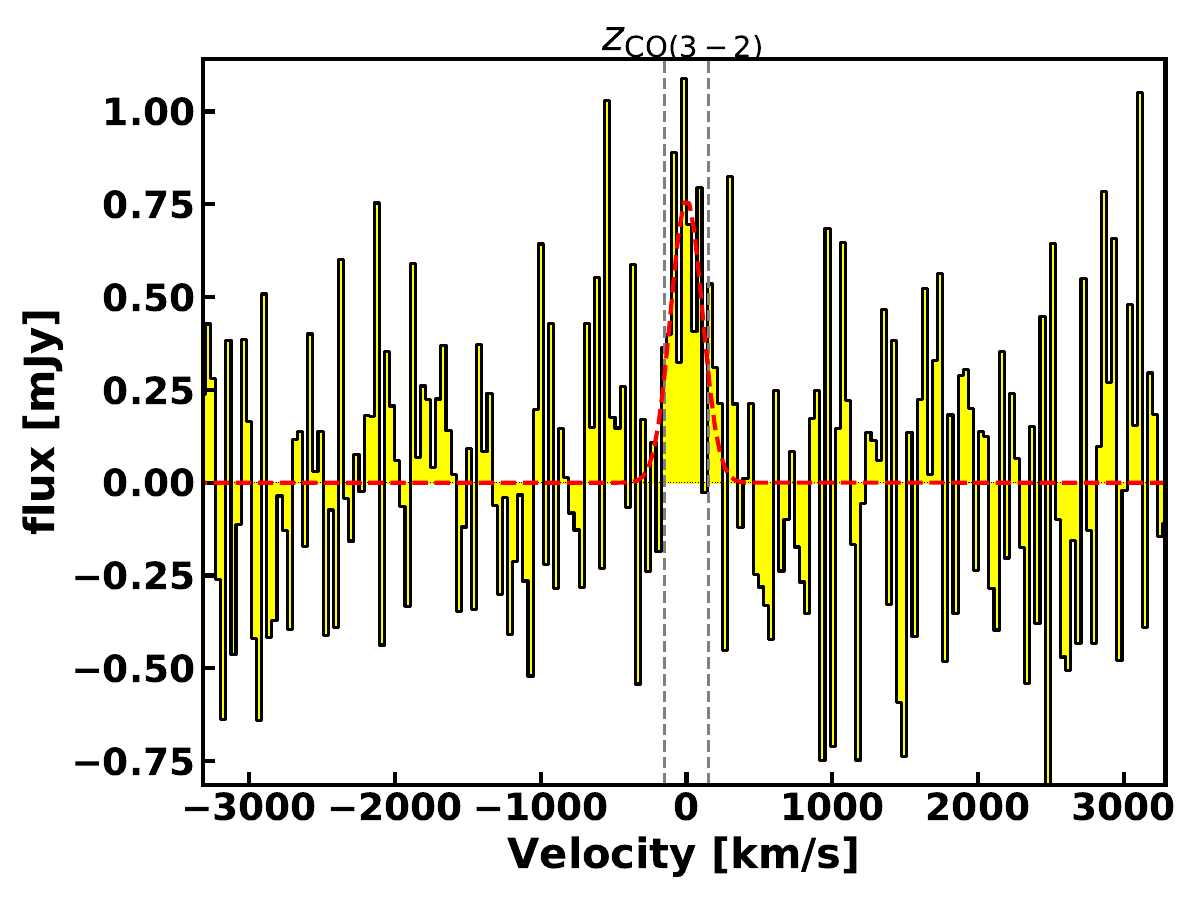}
\includegraphics[width=0.2465\textwidth]{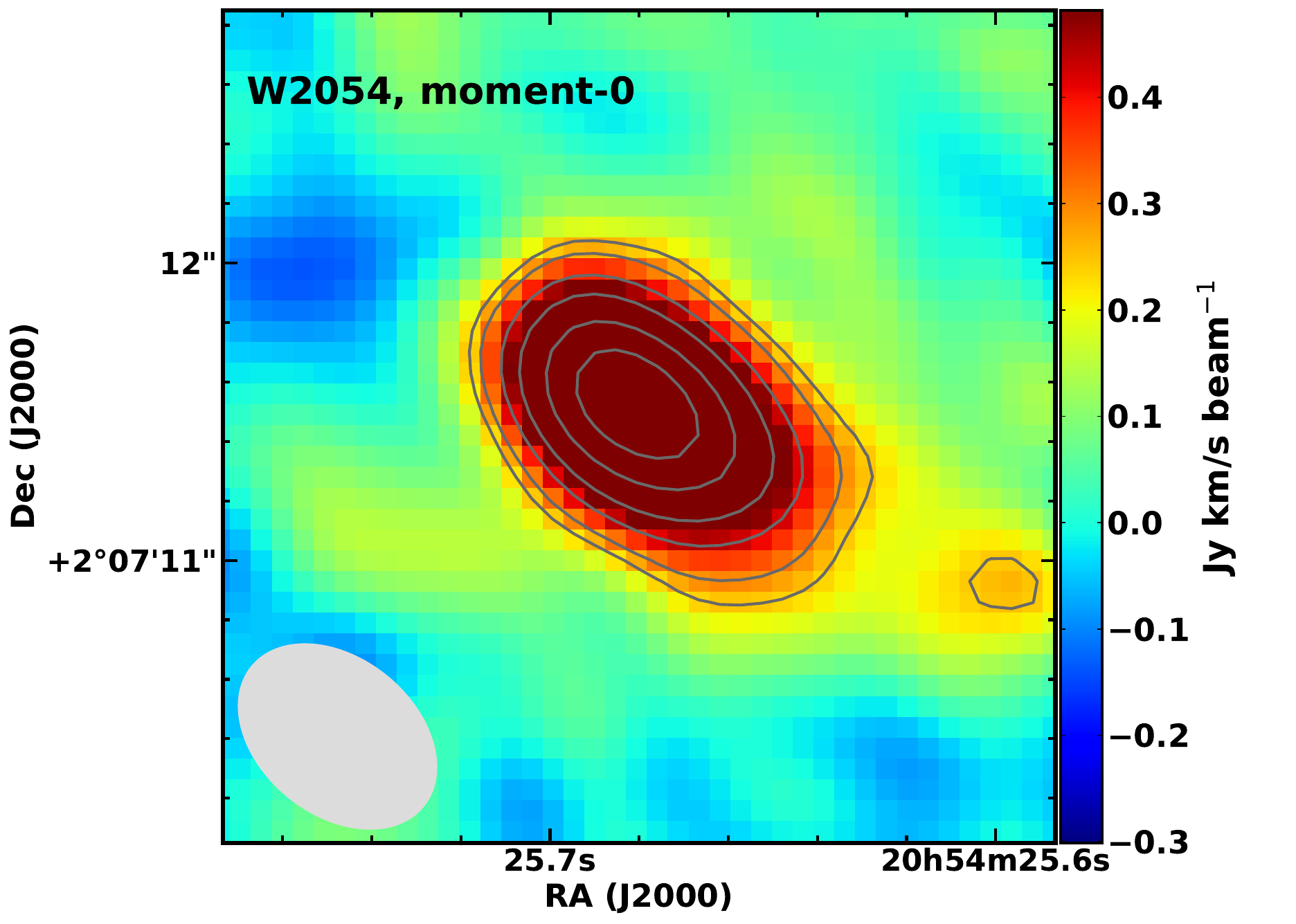}
\includegraphics[width=0.2465\textwidth]{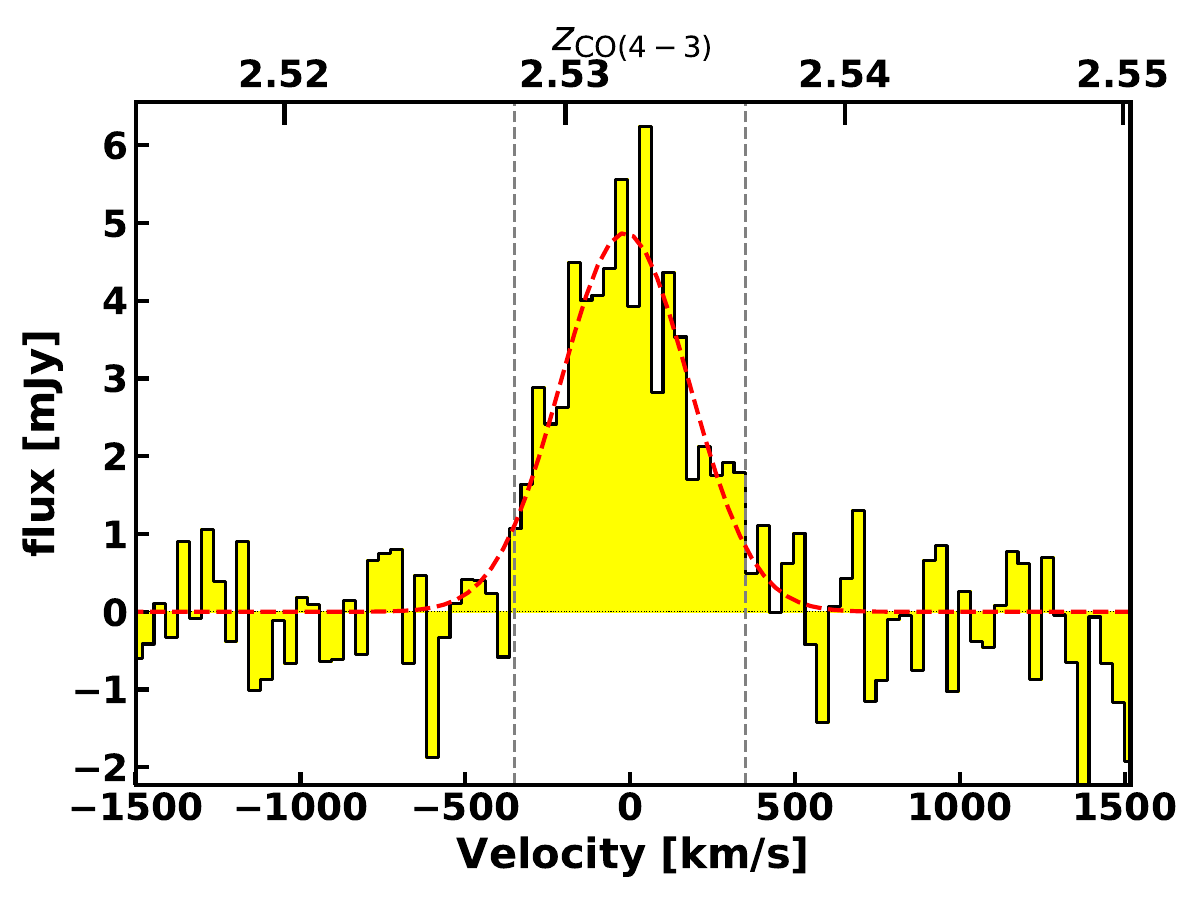}
\includegraphics[width=0.2465\textwidth]{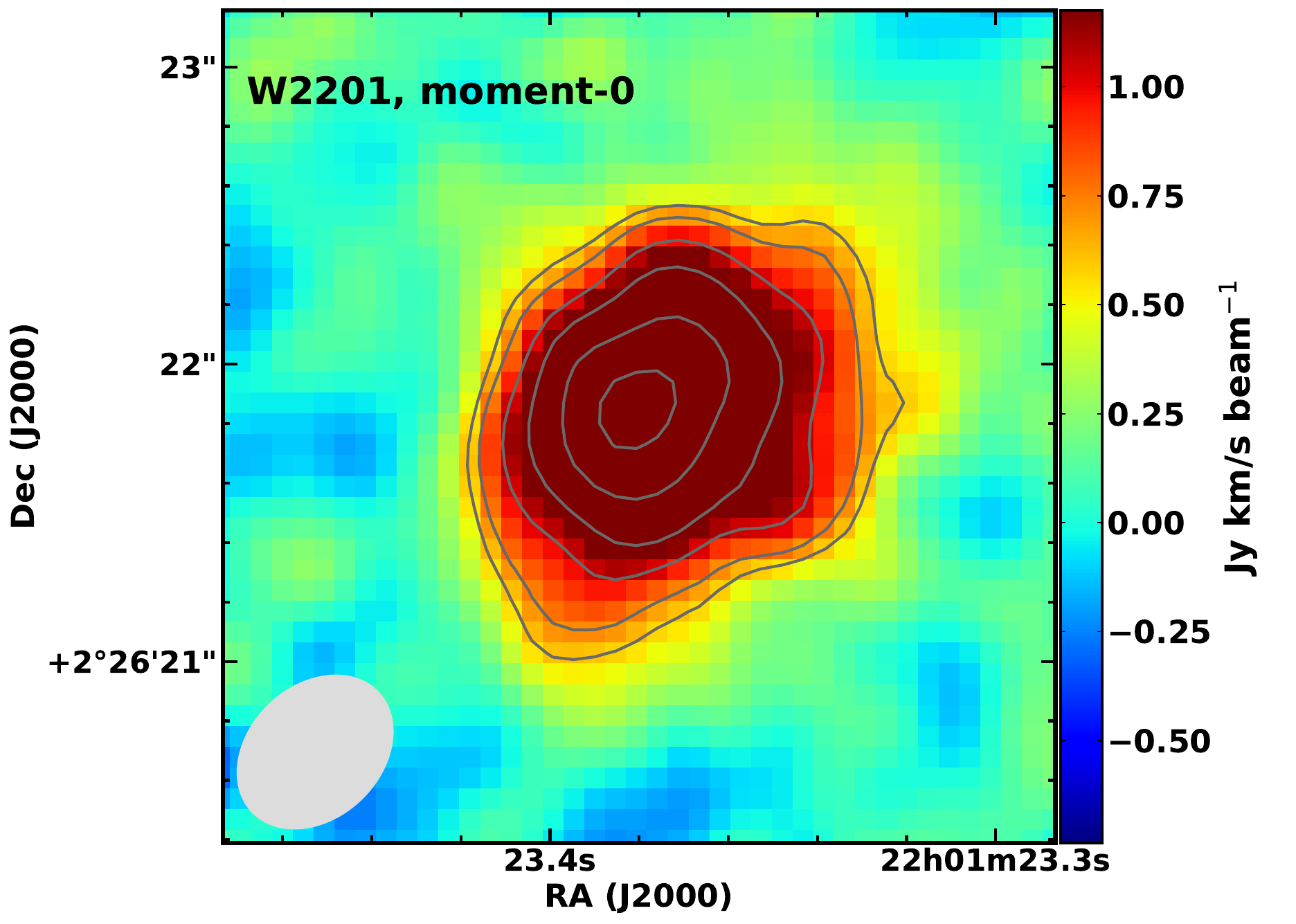}
\includegraphics[width=0.2465\textwidth]{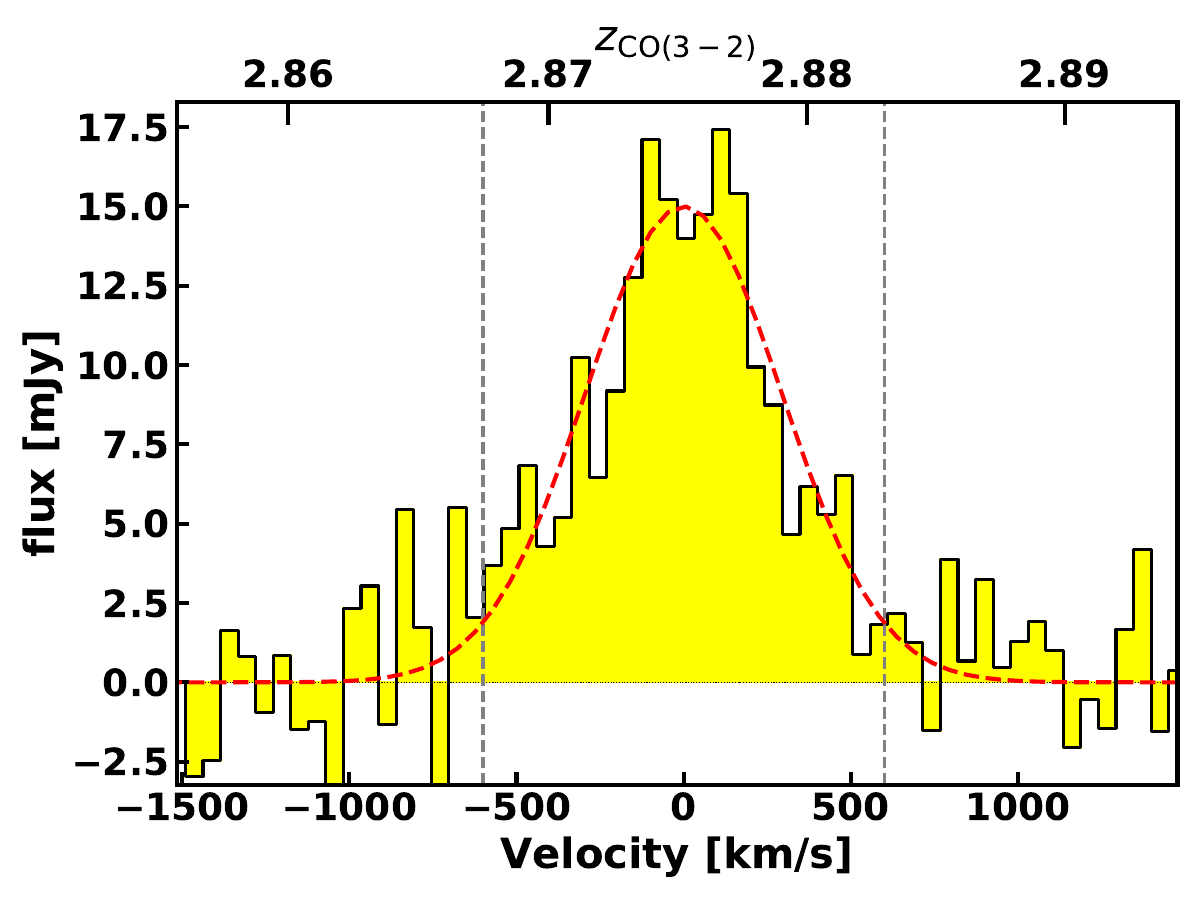}
\includegraphics[width=0.2465\textwidth]{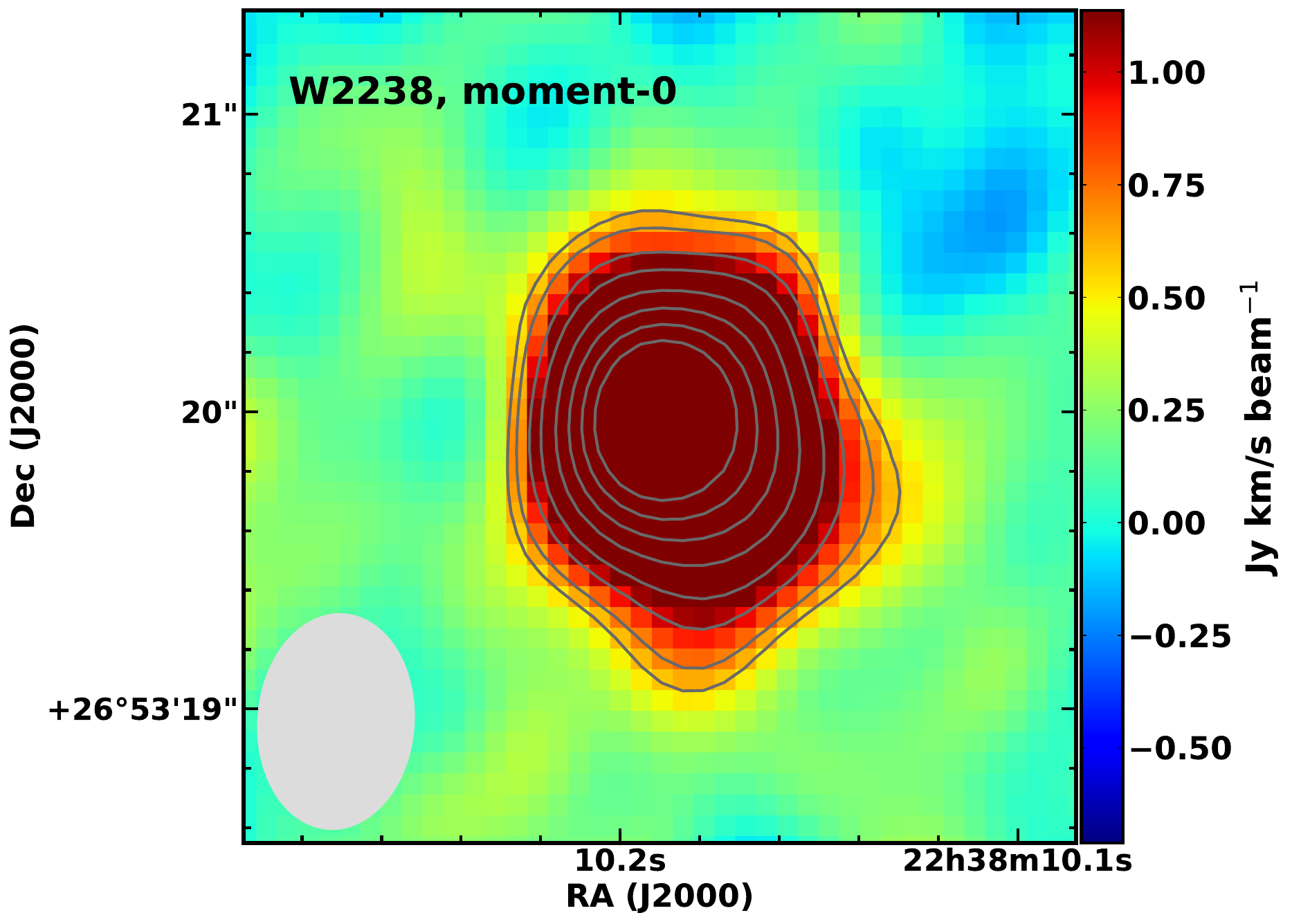}
\includegraphics[width=0.2465\textwidth]{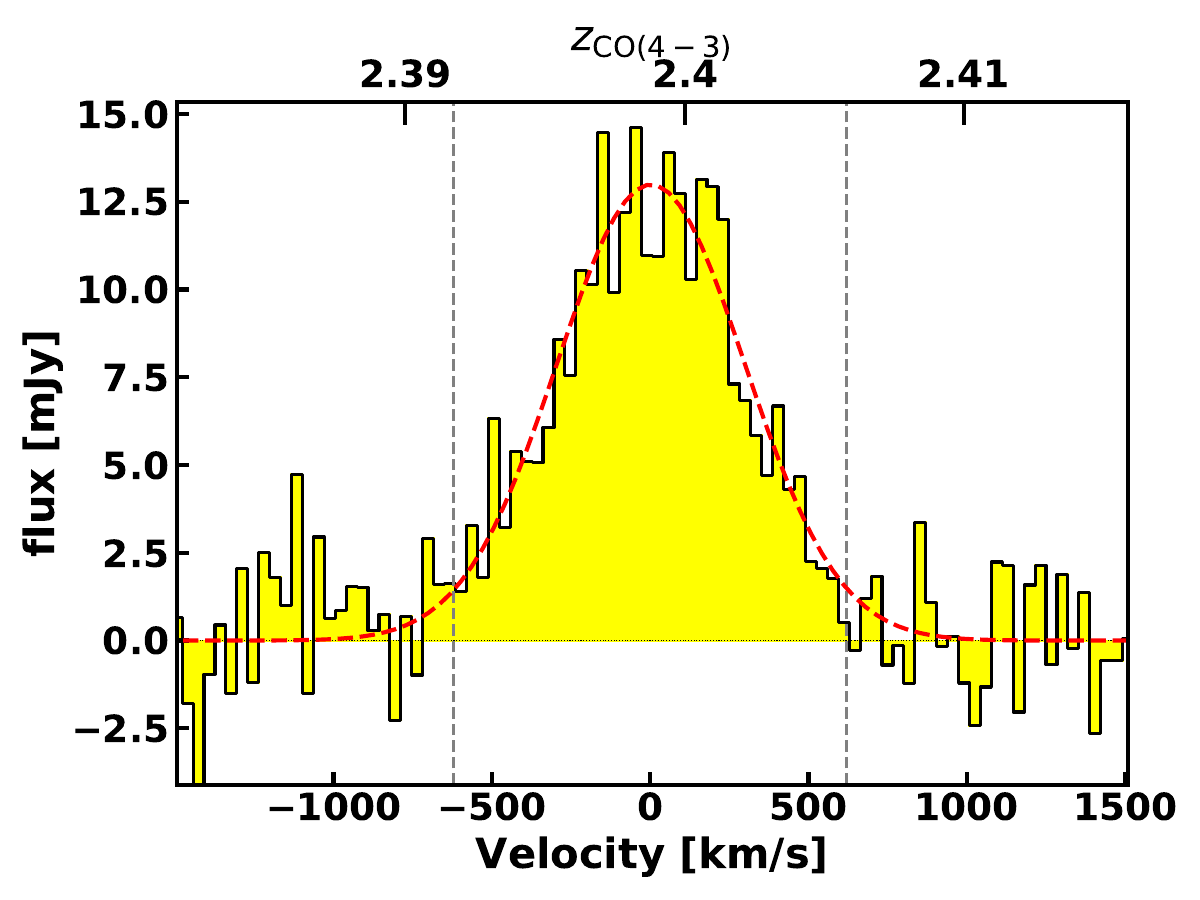}
\includegraphics[width=0.2465\textwidth]{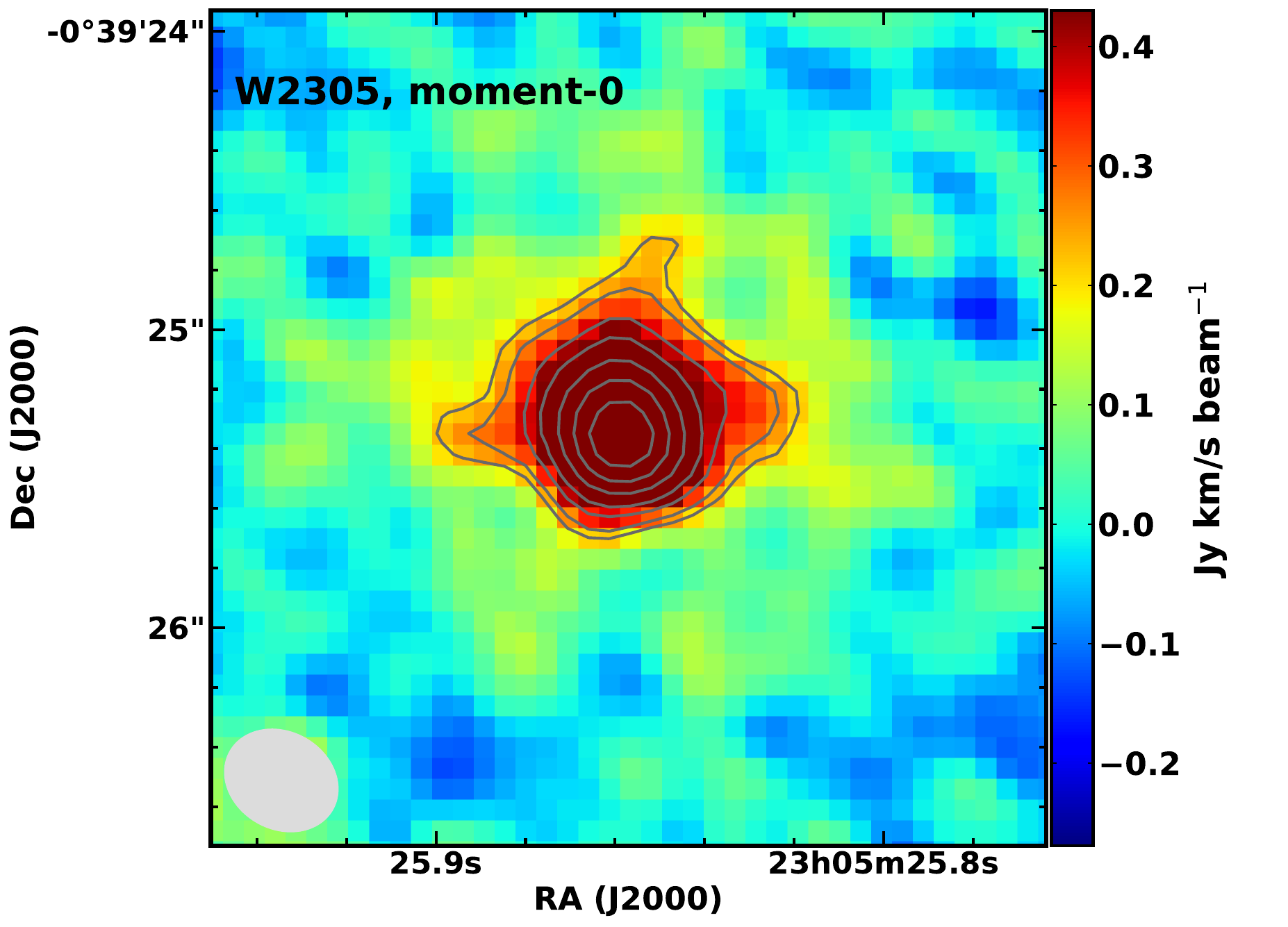}
\includegraphics[width=0.2465\textwidth]{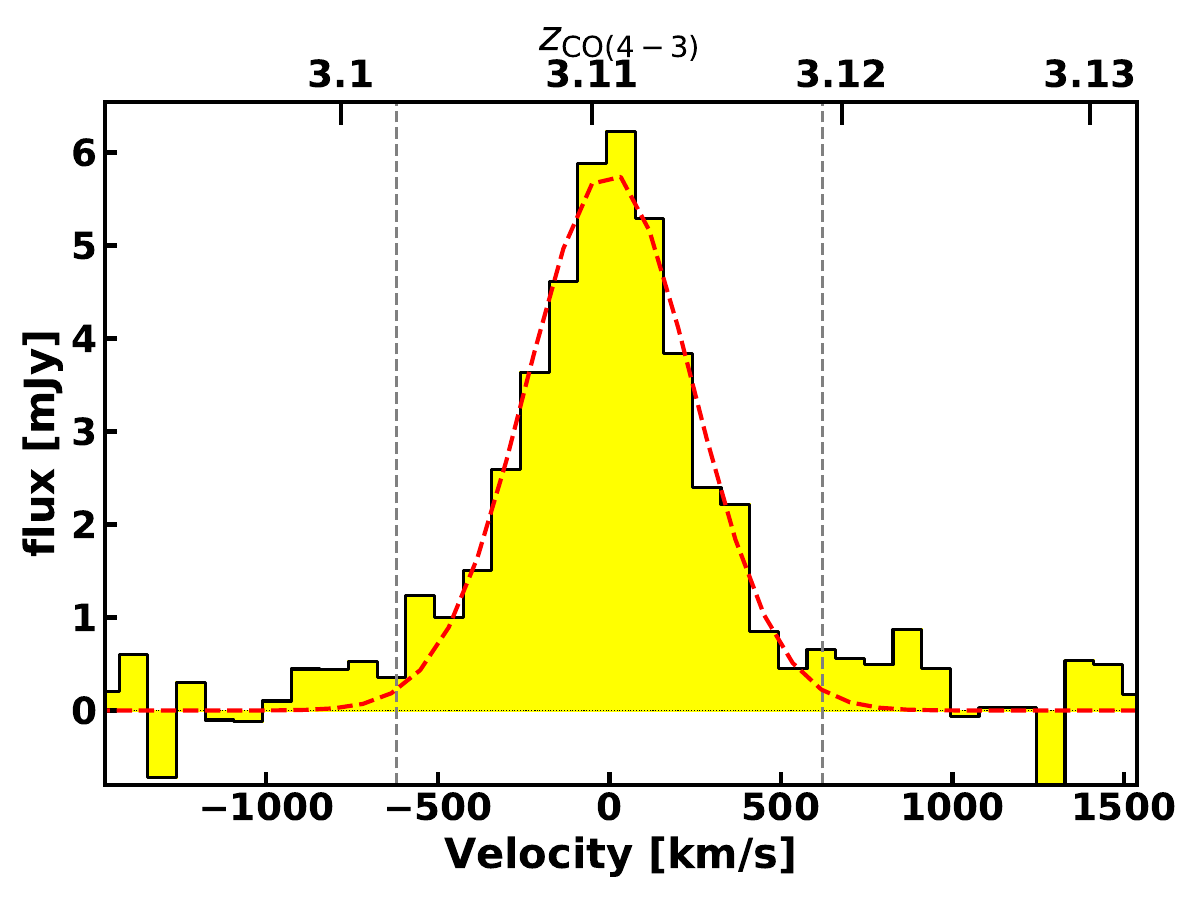}
\caption{The CO emission line moment 0 maps and continuum-subtracted spectra.} 
\label{img:CO}
\end{figure*}

We detect significant CO(3-2) or CO(4-3) emission lines for all but four galaxies in our sample. Among the four galaxies, W0248+2705 and W1814+3412 have a tentative detection, while no emission lines were found for the other two sources. The emission lines were fitted with a single Gaussian
line profile to estimate the redshift, peak flux density, line width, and integrated intensity. The results are given in
Table~\ref{tab:results}. For the four nondetections, 3$\sigma$ upper limits of the peak flux density and integrated intensity (assuming a line width of 300\,km\,s$^{-1}$) are reported in Table~\ref{tab:results}. Galaxy centers were
estimated from the moment-0 maps. Figure~\ref{img:CO} shows the CO line spectra and moment 0 maps of our galaxies. 

The continuum was detected in all but four galaxies. The continuum flux density and size were measured by fitting 2D Gaussian profile to the maps. We note that W0126-0529 has no line
detection but has a clear detection of its continuum. The relevant measurements of continuum are reported in
Table~\ref{tab:results}, where nondetections are listed as $5\sigma$
upper limits.

The line luminosities $L^{'}_{\rm CO}$ in {K\,km\,s$^{-1}$\,pc$^2$} were calculated by using an equation from \citet{solomon2005}: \begin{equation}\label{equ:LCO}
    L^{'}_{\rm CO}=3.25\times10^7\,S_{\rm CO}\,\Delta v\,\nu_{\rm obs}^{-2}\,D_{\rm L}^2\, (1+z)^{-3},
\end{equation}
where $S_{\rm CO} \Delta v$ is the CO line flux in Jy\,km\,s$^{-1}$, $D_{\rm L}$ is the luminosity distance in Mpc, and $\nu_{\rm obs}$ is the observed frequency in GHz. The line emission data for W0149+2305, W0220+0137, and W0410-0913 were collected from \citet{Fan2018b}, and the line emission data for W2246-0526 were obtained from \citet{Diaz-santos2018}. W0126-0529 has been re-observed recently and a redshift of z = 0.832 \citep{Vito2018} has been reported, suggesting that it may be classified as a low-redshift Hot DOG, similar to W1904+4853 in \citet{Li2023}. An uncertain redshift might be responsible for its nondetection in our observations. Thus, we chose to discard this source from our analysis. Utilizing line ratios of $L_{\rm CO(3-2)}^{'}$ / $L_{\rm CO(1-0)}^{'}$ = 0.97 and $L_{\rm CO(4-3)}^{'}$ / $L_{\rm CO(1-0)}^{'}$ = 0.87, as recommended for QSOs \citep{Carilli2013}, we derived the CO(1-0) line luminosities. We note that a intermediate value of $L_{\rm CO(3-2)}^{'}$ / $L_{\rm CO(1-0)}^{'}$ = 0.8 between SMGs and QSOs was adopted in \citet{Banerji2017} and \citet{Fan2019}. However, our results in Section \ref{section4.2} indicate that the bolometric luminosities of our sample are dominated by central AGN emission, favoring the line ratios typical of QSOs. The CO-to-H$_2$ conversion factor, $\alpha_{\rm CO}$, relates the CO(1-0) luminosity to total molecular gas mass. For the Milk Way, we have $\alpha_{\rm CO}\sim\,4.6\,\rm M_\odot$(K\,km\,s$^{-1}$\,pc$^2$)$^{-1}$ \citep{Bolatto2013}. However, for starburst galaxies, $\alpha_{\rm CO}$ is significantly lower, with a value of $0.8\,\rm M_\odot$(K\,km\,s$^{-1}$\,pc$^2$)$^{-1}$ \citep{Carilli2013}. We adopted $\alpha_{\rm CO} = 0.8\,\rm M_\odot$(K\,km\,s$^{-1}$\,pc$^2$)$^{-1}$ for our galaxies, given the starburst nature \citep{Fan2019} and the high merger fraction \citep{Fan2016a} of Hot DOGs. The logarithmic molecular gas mass log $M_{\rm H_2}[\rm M_\odot]$ of our sample ranges from 9.55 to 11.59, with a median value of 10.56. Our estimations of molecular gas mass based on CO line observations is 0.56 dex higher than the values predicted by dust mass assuming a Milky Way dust-to-gas ratio of 0.01 in \citet{Fan2016b}. The line luminosities and molecular gas mass of our sample are listed in Table \ref{table:CO line}

\begin{deluxetable}{lccccc}
\tablecaption{CO Line Luminosities and Molecular Gas Masses.\label{table:CO line}}
\tabletypesize{\scriptsize}
\setlength{\tabcolsep}{2pt}
\tablehead{
\\
\colhead{Source}  &  \colhead{Redshift}  &  \colhead{$L^{'}_{\rm CO(4-3)}$}    &   \colhead{$L^{'}_{\rm CO(3-2)}$}    &   \colhead{$L^{'}_{\rm CO(1-0)}$}    &   \colhead{$M_{\rm H_2}$}\\
\colhead{(1)} & \colhead{(2)} & \colhead{(3)} & \colhead{(4)} & \colhead{(5)} & \colhead{(6)}
}
\startdata
W0134-2922  &  3.057  & $1.48\pm0.24$   & ..    &  $1.7\pm0.28$ &   $1.36\pm0.22$\\
W0149+2350$^a$  &  3.23  &  $2.40\pm0.50$    &   ..  &   $2.76\pm0.57$  &   $2.21\pm0.46$\\
W0220+0137$^a$  &  3.136  & $3.39\pm0.65$   &   ..  &   $3.9\pm0.75$    &   $3.12\pm0.60$\\
W0248+2705  &  2.183  & $1.12\pm0.40$   &   ..  &   $1.29\pm0.46$   &   $1.03\pm0.37$\\
W0410-0913$^a$  &  3.63  &  $17.90\pm1.50$  &   ..  &   $20.57\pm1.72$  &   $16.46\pm1.38$\\
W0533-3401  &  2.902  & ..  &  $12.50\pm1.70$    &   $12.88\pm1.75$ &   $10.30\pm1.40$\\
W0615-5716  &  3.346  & $1.30\pm0.27$    &   ..  &   $1.49\pm0.31$  &   $1.19\pm0.25$\\
W1248-2154  &  3.323  & $0.62\pm0.15$   &   ..  &   $0.71\pm0.18$   &   $0.57\pm0.14$\\
W1603+2745  &  2.654  & .. &   $7.65\pm0.96$    &   $7.89\pm0.99$   &   $6.31\pm0.79$\\
W1814+3412  &  2.457  & $0.38\pm0.15$   &   ..  &   $0.44\pm0.17$   &   $0.35\pm0.14$\\
W2054+0207  &  2.532  & $4.63\pm0.45$   &   ..  &   $5.32\pm0.52$   &   $4.26\pm0.42$\\
W2201+0226  &  2.875  & ..  &   $47.4\pm4.20$   &   $48.87\pm4.33$  &   $39.10\pm3.46$\\
W2210-3507  &  2.814  & ..  &   $<1.40$  &   $<1.44$    &   $<1.15$\\
W2238+2653  &  2.399  & $17.20\pm1.00$   &   ..  &   $19.77\pm1.15$  &  $15.82\pm0.92$\\
W2246-0526$^b$  &  4.601  & ..  &   ..  &   $8.98\pm1.90$   &   $7.18\pm1.52$\\
W2305-0039  &  3.111  & $9.66\pm0.59$   &   ..  &   $11.10\pm0.68$  &   $8.88\pm0.54$\\
\enddata
\tablecomments{(1): Source name. $^a$ From \citet{Fan2018b}. $^b$ From \citet{Diaz-santos2018}. (2): The spectroscopic redshift. (3）and (4): Observed CO(4-3) or CO(3-2) line luminosity (10$^{10}$\,K\,km\,s$^{-1}$\,pc$^2$) calculated from line flux $I_{\rm CO}$ in table \ref{tab:results}. (5): CO(1-0) line luminosity (10$^{10}$\,K\,km\,s$^{-1}$\,pc$^2$) adopting $L_{\rm CO(3-2)}^{'}$/$L_{\rm CO(1-0)}^{'}$ = 0.97 and $L_{\rm CO(4-3)}^{'}$/$L_{\rm CO(1-0)}^{'}$ = 0.87. (6): Molecular gas mass (10$^{10}\rm M_\odot]$) assuming $\alpha_{\rm CO}\sim 0.8\,\rm M_\odot$(K\,km\,s$^{-1}$\,pc$^2$)$^{-1}$.}
\end{deluxetable}

\subsection{Results of the SED Fitting and Dust Properties} \label{section4.2}
\begin{figure*}
\centering
\fig{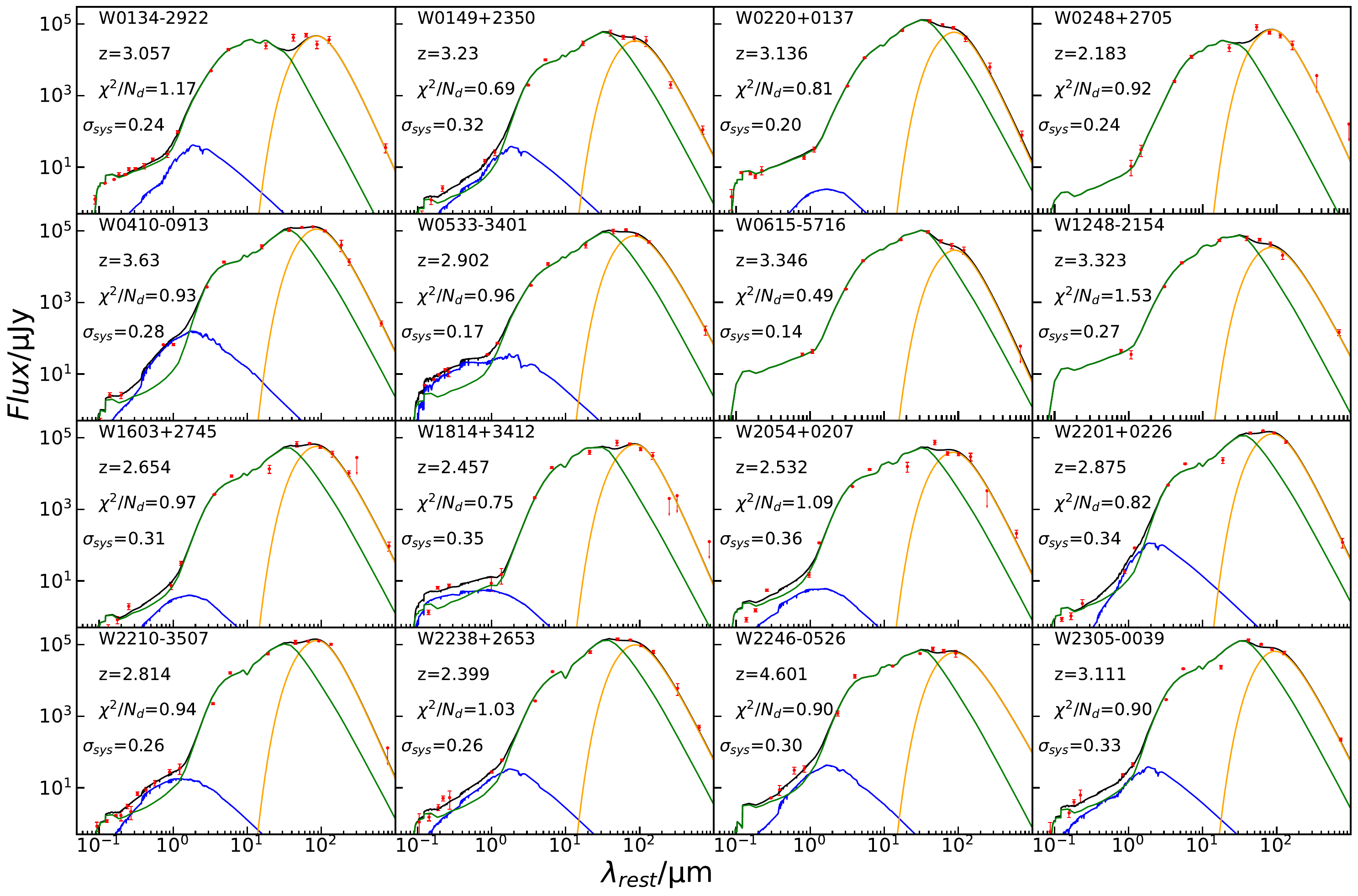}{1\textwidth}{}
\caption{Best-fit model SEDs by $T_{\rm dust}$-constrained fitting for 16 Hot DOGs in our sample. The source ID, redshift, $\chi^2/N_d$ ($N_d$ is the photometric data points), and the systematical errors \citep[see][]{HanY2014a} are shown in each panel. The red points with error bars represent the observed photometric data, and those with downward arrows mark flux density upper limits. The solid blue, orange, and green lines represent, respectively, the stellar, cold dust, and AGN components. The solid black line represents the total SED.}
\label{figure:SED}
\end{figure*}

The best-fit SEDs are shown in Figure \ref{figure:SED}. Thanks to high AGN obscuration, the host galaxies of our Hot DOGs sample are easily observable, and the stellar emission can be separated out so that we can estimate the physical properties of the host galaxies. We adopted 
the median of the posterior probability distribution of each parameter as the fiducial value, and the uncertainties are reported as the 68\% confidence intervals around the fiducial values. The derived properties are listed in Table \ref{table:SED fitting results}. Based on our UV-millimeter three-component SED modeling described in Section \ref{section3}, the stellar population parameters, including stellar masses and SFRs, as well as the cold dust properties from the cold dust component and AGN luminosities have been obtained. 

The cold dust infrared luminosity $L_{\rm IR}$ was calculated by integrating the cold dust component from 8 to 1000 $\mu$m. Our Hot DOGs exhibit $L_{\rm IR}\,\sim\,10^{13}\,L_\odot$. The estimated dust temperatures range from 42K to 48K, with a median of 45 K, consistent with $T_{\rm dust}$-$L_{\rm IR}$ relation of SMGs \citep[e.g.,][]{Magnelli2012,Roseboom2012}. The parameter $\beta$ denotes the power-law index of the optical depth, with $\tau_\lambda$= $(\frac{\lambda_0}{\lambda})^{\beta}$. It ranges from 2.0 to 2.8, with a median of 2.5. The measured $T_{\rm dust}$ and $\beta$ of Hot DOGs is similar to the most luminous quasar at z = 6.327 \citep{Tripodi2023}. A relatively high $\beta$ value indicates optically thick dust in IR bands, which has also been reported in other compact starburst galaxies \citep[e.g.,][]{Scoville2017}. With the cold dust temperature $T_{\rm dust}$ and the emissivity index $\beta$, we derived the dust mass with the formula:
\begin{equation}\label{equ:mdust}
    M_{\rm dust}=\frac{D^2_{\rm L}}{(1+z)}\times\frac{S_{\nu_{\rm obs}}}{\kappa_{\nu_{\rm rest}} B(\nu_{\rm rest},T_{\rm dust})},
\end{equation}
where $D_{\rm L}$ is the luminosity distance, $S_{\nu_{\rm obs}}$ is the flux density at observed 
frequency $\nu_{\rm obs}$, $\kappa_{\nu_{\rm rest}}=\kappa_0(\nu/\nu_0)^\beta$ is the absorption 
coefficient at the corresponding rest-frame frequency, and $B(\nu_{\rm rest},T_{\rm dust})$ is the Planck 
function per unit frequency at temperature $T_{\rm dust}$. We adopted $\kappa_{850\mu m}$ = 3.8 cm$^2\,\rm g^{-1}$ by following  
\citet{Wu2014} and \citet{Fan2016b,Fan2019}. The estimated dust masses of our sample are consistent with previous 
IR SED decomposition results in \citet{Fan2016b}, with a sample median of $8.0\,\times\,10^{7}\rm M_\odot$. We note that $\beta$ was fixed to 1.6 in \citet{Fan2016b} to avoid degeneracy. The uncertainties in the dust masses shown in Table \ref{table:SED fitting results} will be larger if we consider the adopted $\kappa_{850\mu m}$ value which can vary from $\sim$0.4 to $\sim$11 cm$^2\,\rm g^{-1}$ \citep[e.g.,][]{james2002,draine2003,dunne2003,Siebenmorgen2014}. All these dust properties are listed in Table \ref{table:SED fitting results}.

Together with the CO measurements, these parameters allow us to derive the molecular gas fractions and star formation efficiencies (Table \ref{table:properties}).

\begin{deluxetable*}{lcccccccccc}
\tablecaption{Summary of the results of SED fitting. \label{table:SED fitting results}}
\tabletypesize{\scriptsize}
\tablehead{
\\
\colhead{Source}    &  \colhead{$T_{\rm dust}$} & \colhead{$T_{\rm dust,0}$} & \colhead{$\beta$}  &  \colhead{Log($L_{\rm AGN}$)}   &  \colhead{Log($L_{\rm IR}$)} &  \colhead{Log($L_{\rm IR,0}$)}   &  \colhead{Log($M_{\rm dust}$)}     &  \colhead{Log($\rm {SFR}$)}  &  \colhead{Log($M_{\rm *}$)} & \colhead{$\delta_{\rm GDR}$}\\
        &   \colhead{(K)} &   \colhead{(K)} &    &   \colhead{Log($\rm L_\odot$)} &   \colhead{Log($\rm L_\odot$)}    &   \colhead{Log$(\rm L_\odot)$}    &   \colhead{Log$(\rm M_\odot)$}      &   \colhead{Log$(\rm M_\odot/\rm yr)$}    &   \colhead{Log$(\rm M_\odot)$} & \\
\colhead{(1)} & \colhead{(2)} & \colhead{(3)} & \colhead{(4)} & \colhead{(5)} & \colhead{(6)} & \colhead{(7)} & \colhead{(8)} & \colhead{(9)}  & \colhead{(10)}  & \colhead{(11)}
}
\startdata
W0134-2922$^b$  &  48.03$_{-2.34}^{+1.33}$ &  72.73$_{-9.68}^{+9.75}$  &  2.78$_{-0.16}^{+0.12}$  &  13.98$_{-0.02}^{+0.02}$  &  12.97$_{-0.02}^{+0.01}$  &  13.18$_{-0.07}^{+0.06}$  &  7.40$_{-0.10}^{+0.15}$  &  2.91$_{-0.16}^{+0.13}$  &  10.61$_{-0.35}^{+0.45}$  &  453$_{-187}^{+101}$\\
W0149+2350  &  44.64$_{-4.56}^{+3.29}$ & 84.04$_{-14.72}^{+9.40}$  &  2.42$_{-0.23}^{+0.23}$  &  13.92$_{-0.03}^{+0.02}$  &  12.92$_{-0.05}^{+0.05}$  &  13.31$_{-0.13}^{+0.07}$  &  7.81$_{-0.17}^{+0.18}$  &  2.77$_{-0.21}^{+0.17}$  &  10.82$_{-0.41}^{+0.35}$  &  339$_{-196}^{+133}$\\
W0220+0137$^b$  &  45.33$_{-3.78}^{+2.97}$ & 63.58$_{-11.39}^{+12.60}$  &  2.72$_{-0.16}^{+0.15}$  &  14.15$_{-0.02}^{+0.02}$  &  13.09$_{-0.05}^{+0.05}$  &  13.31$_{-0.13}^{+0.13}$  &  7.68$_{-0.12}^{+0.14}$  &  3.06$_{-0.15}^{+0.11}$  &  10.55$_{-0.23}^{+0.38}$  &  638$_{-286}^{+202}$\\
W0248+2705$^a$  &  47.44$_{-3.50}^{+1.76}$  &  69.16$_{-7.67}^{+6.67}$  &  2.47$_{-0.40}^{+0.33}$  &  13.51$_{-0.02}^{+0.02}$  &  12.89$_{-0.05}^{+0.02}$  &  13.12$_{-0.06}^{+0.05}$  &  7.61$_{-0.26}^{+0.34}$  &  2.78$_{-0.05}^{+0.02}$  & \nodata  &  250$_{-314}^{+145}$\\
W0410-0913  &  47.35$_{-2.57}^{+1.78}$  &  74.62$_{-8.83}^{+8.30}$  &  2.63$_{-0.18}^{+0.16}$  &  14.21$_{-0.03}^{+0.04}$  &  13.51$_{-0.03}^{+0.03}$  &  13.80$_{-0.08}^{+0.05}$  &  8.09$_{-0.12}^{+0.14}$  &  3.08$_{-0.04}^{+0.13}$  &  12.08$_{-0.25}^{+0.10}$  &  1325$_{-549}^{+346}$\\
W0533-3401  &  46.52$_{-4.07}^{+2.37}$  &  79.21$_{-8.45}^{+7.05}$  &  2.26$_{-0.19}^{+0.21}$  &  13.98$_{-0.02}^{+0.02}$  &  13.14$_{-0.06}^{+0.04}$  &  13.55$_{-0.07}^{+0.05}$  &  8.08$_{-0.16}^{+0.14}$  &  3.16$_{-0.17}^{+0.12}$  &  10.65$_{-0.23}^{+0.34}$  &  845$_{-369}^{+297}$\\
W0615-5716$^a$  &  45.34$_{-3.88}^{+3.11}$  &  73.38$_{-17.19}^{+15.42}$  &  2.75$_{-0.18}^{+0.15}$  &  14.21$_{-0.01}^{+0.01}$  &  12.83$_{-0.04}^{+0.05}$  &  13.14$_{-0.20}^{+0.15}$  &  7.43$_{-0.16}^{+0.15}$  &  2.72$_{-0.04}^{+0.05}$  &  \nodata  &  440$_{-215}^{+168}$\\
W1248-2154$^a$  &  45.37$_{-5.02}^{+3.15}$  &  74.59$_{-12.23}^{+11.83}$  &  2.13$_{-0.19}^{+0.21}$  &  14.14$_{-0.03}^{+0.03}$  &  12.91$_{-0.07}^{+0.05}$  &  13.22$_{-0.11}^{+0.09}$  &  8.00$_{-0.15}^{+0.16}$  &  2.80$_{-0.07}^{+0.05}$  &  \nodata  &  56$_{-29}^{+21}$\\
W1603+2745$^b$  &  44.87$_{-5.02}^{+3.32}$  &  67.00$_{-7.66}^{+7.02}$  &  2.35$_{-0.20}^{+0.21}$  &  13.65$_{-0.03}^{+0.03}$  &  12.98$_{-0.08}^{+0.06}$  &  13.26$_{-0.06}^{+0.05}$  &  7.91$_{-0.15}^{+0.16}$  &  2.91$_{-0.19}^{+0.14}$  &  10.58$_{-0.31}^{+0.46}$  &  762$_{-359}^{+247}$\\
W1814+3412  &  46.59$_{-3.61}^{+2.32}$  &  73.43$_{-11.05}^{+10.89}$  &  2.75$_{-0.24}^{+0.16}$  &  13.73$_{-0.02}^{+0.02}$  &  12.93$_{-0.05}^{+0.03}$  &  13.20$_{-0.09}^{+0.08}$  &  7.45$_{-0.16}^{+0.23}$  &  2.92$_{-0.13}^{+0.10}$  &  10.34$_{-0.19}^{+0.31}$  &  122$_{-98}^{+60}$\\
W2054+0207$^b$  &  41.98$_{-5.84}^{+5.23}$  &  79.23$_{-8.57}^{+8.39}$  &  2.53$_{-0.27}^{+0.25}$  &  13.72$_{-0.03}^{+0.03}$  &  12.71$_{-0.09}^{+0.09}$  &  13.18$_{-0.07}^{+0.06}$  &  7.64$_{-0.19}^{+0.21}$  &  2.59$_{-0.24}^{+0.19}$  &  10.54$_{-0.46}^{+0.52}$  &  974$_{-642}^{+366}$\\
W2201+0226  &  45.85$_{-3.79}^{+2.75}$  &  72.06$_{-6.73}^{+5.56}$  &  2.70$_{-0.23}^{+0.16}$  &  14.03$_{-0.03}^{+0.03}$  &  13.39$_{-0.05}^{+0.05}$  &  13.69$_{-0.05}^{+0.03}$  &  8.00$_{-0.16}^{+0.19}$  &  3.26$_{-0.23}^{+0.18}$  &  11.24$_{-0.37}^{+0.39}$  &  3886$_{-2217}^{+1278}$\\
W2210-3507  &  45.32$_{-3.88}^{+2.84}$  &  60.33$_{-9.12}^{+9.25}$  &  2.78$_{-0.19}^{+0.13}$  &  13.98$_{-0.02}^{+0.02}$  &  13.35$_{-0.04}^{+0.03}$  &  13.52$_{-0.09}^{+0.07}$  &  7.92$_{-0.14}^{+0.17}$  &  3.23$_{-0.18}^{+0.17}$  &  11.14$_{-0.38}^{+0.38}$  &  162$_{-84}^{+53}$\\
W2238+2653  &  44.76$_{-5.48}^{+3.59}$  &  72.92$_{-8.90}^{+9.28}$  &  2.41$_{-0.19}^{+0.22}$  &  13.90$_{-0.02}^{+0.02}$  &  13.15$_{-0.08}^{+0.06}$  &  13.50$_{-0.08}^{+0.07}$  &  8.04$_{-0.14}^{+0.14}$  &  3.07$_{-0.20}^{+0.16}$  &  10.76$_{-0.31}^{+0.47}$  &  1433$_{-586}^{+422}$\\
W2246-0526  &  44.31$_{-5.24}^{+3.58}$  &  91.56$_{-12.21}^{+5.84}$  &  1.96$_{-0.56}^{+0.64}$  &  14.33$_{-0.04}^{+0.03}$  &  13.41$_{-0.05}^{+0.05}$  &  13.78$_{-0.08}^{+0.03}$  &  8.70$_{-0.53}^{+0.52}$  &  3.24$_{-0.22}^{+0.20}$  &  11.40$_{-0.50}^{+0.34}$  &  141$_{-330}^{+104}$\\
W2305-0039  &  42.75$_{-6.28}^{+4.66}$  &  78.73$_{-10.15}^{+7.18}$  &  2.28$_{-0.22}^{+0.23}$  &  14.14$_{-0.02}^{+0.02}$  &  13.17$_{-0.09}^{+0.07}$  &  13.63$_{-0.08}^{+0.05}$  &  8.25$_{-0.14}^{+0.18}$  &  3.06$_{-0.20}^{+0.17}$  &  10.89$_{-0.39}^{+0.46}$  &  494$_{-259}^{+142}$\\
\enddata
\tablecomments{Median and 16th$-$84th quartile ranges of the parameter posterior probability distribution. (1): Source name. $^a$ Sources without UV-optical data. \edit1{$^b$ Sources identified as BHDs, as discussed in Section \ref{sec:5.5}}. (2) Cold dust temperature. (3) Cold dust temperature, but for the $T_{\rm dust}$-unconstrained fitting. (4) Dust emissivity index in the gray body function. (5) AGN bolometric luminosity by integrating the AGN component SED. (6): Host galaxy infrared luminosity  by integrating the cold dust component SED. (7): Host galaxy infrared luminosity, but for the $T_{\rm dust}$-unconstrained fitting. (8): Dust mass. (9) Star formation rate. (10) Stellar mass. \edit1{For BHDs, their stellar mass may have larger uncertainties. } (11) Gas-to-dust ratio}
\end{deluxetable*}

\begin{deluxetable*}{lcccccccc}
\centering
\tablecaption{Summary of the physical properties. \label{table:properties}}
\tabletypesize{\scriptsize}
\tablehead{
\\
\colhead{statistics value} & \colhead{$\delta_{\rm GDR}$} & $f_{\rm gas}$  & $\Delta_{\rm MS}$ & $t_{\rm depl}$ & SFE & M$_{\rm BH}$ & $M_{\rm BH}$/$M_\star$ & ${\dot{M}}_{\rm BH}$\\
 & & & & [Myr] & [K km $\rm s^{-1}$ $\rm pc^{-2}]$ & [10$^{9}\rm M_\odot]$ & & $\rm M_\odot\,\rm yr^{-1}$\\
\colhead{(1)} & \colhead{(2)} & \colhead{(3)} & \colhead{(4)} & \colhead{(5)} & \colhead{(6)} & \colhead{(7)} & \colhead{(8)}  & \colhead{(9)}
 }
\startdata
max & 3886  &   0.73 & 15 & 222 & 1958 & 6.6 & 0.122 & 147\\
median, scatter & $474_{-324}^{+711}$ & $0.33_{-0.17}^{+0.33}$ & 6.12$^{+5.1}_{-2.9}$ & 39$_{-28}^{+85}$ & 297$_{-195}^{+659}$ & 3.0$_{-1.3}^{+1.8}$  & 0.042$_{-0.021}^{+0.029}$  & 65$_{-29}^{+39}$\\
min & 56 & 0.09 & 0.62 & 4 & 51 & 1.0 & 0.004 & 22\\
\enddata
\tablecomments{(1): Sample maximum, median and 16th$-$84th quartile range and minimum.（2) Gas-to-dust ratio. (3) Molecular gas fraction. (4): SFR offset against main-sequence (MS) galaxies (5): Gas depletion timescale. (6) SFE. (7): Central black hole mass assuming $\lambda_{\rm Edd}$ = 1.0. (8) Black hole to stellar mass ratio. (9): Black hole growth rate.}
\end{deluxetable*}

\section{Discussion}\label{section5}

\subsection{Gas-to-dust ratio}

\begin{figure}[!htb]
\centering
\resizebox{1.0\hsize}{!}{\includegraphics{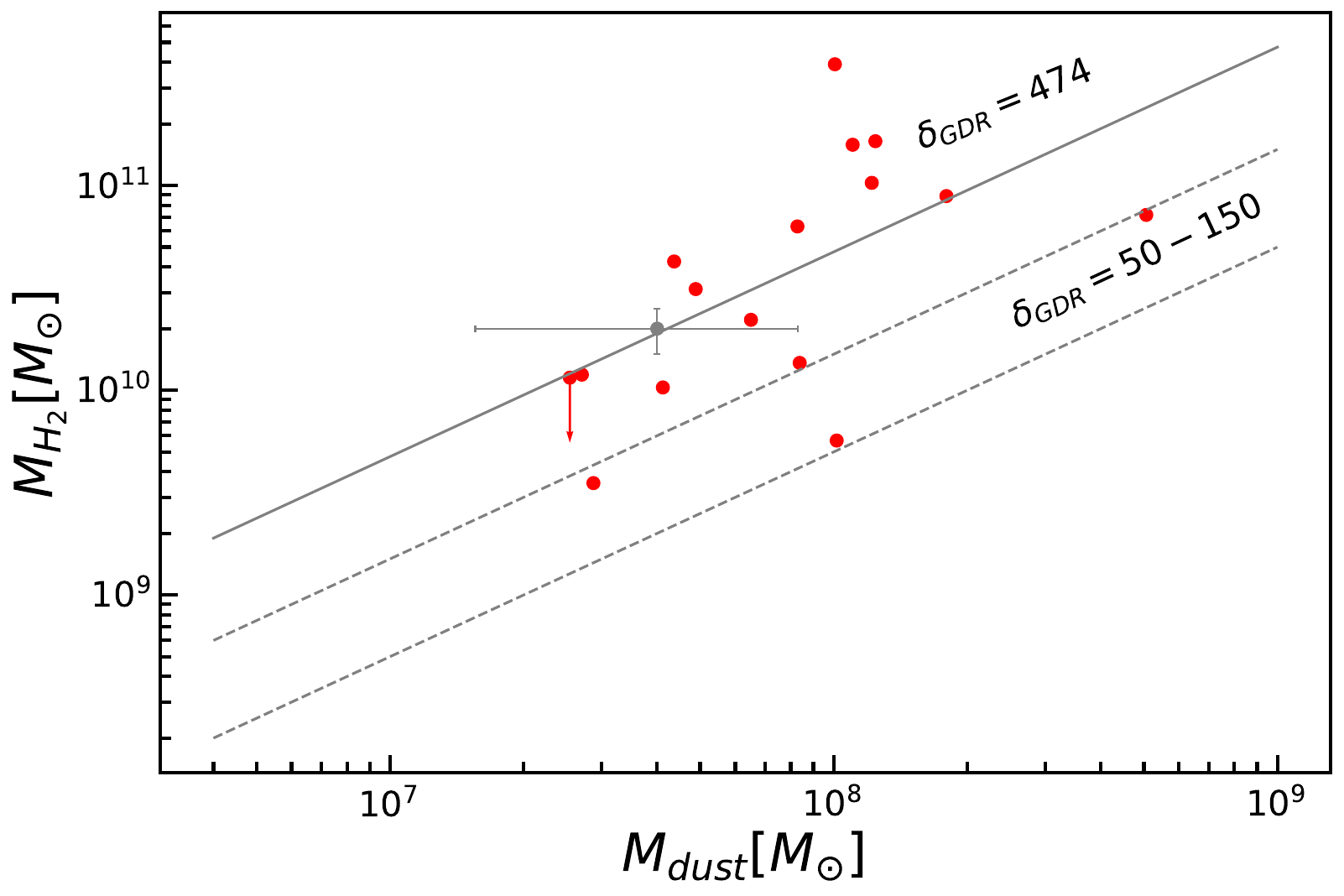}}
\caption{Molecular gas mass $M_{\rm H_2}$ vs. cold dust mass $M_{\rm dust}$. The red points represent our sample values, while the downward arrows mark the upper limit molecular gas mass for W2210-3507. The typical uncertainties are shown as a gray point with an error bar. The two dashed lines represent a gas-to-dust ratio $\delta_{\rm GDR}$ value of 50-150, which is the typical value derived for the Milky Way \citep{Jenkins2004}, 
local star forming galaxies \citep{Draine2007,Remy2014} and high-redshift SMGs \citep{Magnelli2012,Miettinen2017}. The solid line represents a gas-to-dust ratio $\delta_{\rm GDR}$ value of 521, which is typical for our sample based on SED analysis and ALMA observations.}
\label{fig:dust}
\end{figure}

The estimated molecular gas mass $M_{\rm H_2}$ is plotted as a function of $M_{\rm dust}$ in Figure \ref{fig:dust}. The two dashed lines represent gas-to-dust ratios $\delta_{\rm GDR}\,=$ 50 and 150, which cover the typical values derived for the Milky Way \citep{Jenkins2004}, 
local star-forming galaxies \citep{Draine2007,Remy2014}, and high-redshift SMGs \citep{Magnelli2012,Miettinen2017}. Most of our Hot DOGs exhibit a high $\delta_{\rm GDR}$ value $\sim$ $474_{-324}^{+711}$, with a median uncertainty of 43\% (uncertainties propagated from $M_{\rm H_2}$ and $M_{\rm dust}$). \citet{Bischetti2021} found a median $\delta_{\rm GDR}$ value of 180 %(\tred{it is NOT similiar to 1000 mentioned above})
for their nine hyperluminous, Type I quasars at $z \sim 2-4$, slightly higher than the typical values, and they attributed this to an increasing $\delta_{\rm GDR}$ with redshift \citep[e.g.,][]{Miettinen2017}. We note that our sample is selected to have Herschel PACS and SPIRE observations, and have either SPIRE 500 $\mu$m or SCUBA-2 850 $\mu$m detection, and therefore may be biased toward the most intense starbursting systems, where the supernova-shock-driven dust heating and destruction may be more significant \citep{Jones2004}. The dust mass decreases with increasing dust temperature \citep[][and references there in]{Fan2016b}. The estimated $T_{\rm dust}\sim$ 45 K may trace a warmer dust component associated with photodissociation regions by starbursts, instead of the diffuse ISM with temperature below 30 K which represents the bulk of the dust mass \citep{Draine2007b,Liang2019,Sommovigo2020,Pozzi2021}. The dust mass estimated from a simplistic graybody model may be underestimated by up to a factor of 3 compared to that derived by the dust models \citep[for example, ][]{Draine2007b} which include more parameters and adequately describe the multicomponent dust properties \citep[][and references therein]{Conroy2013}. When we take a dust temperature of 30 K we found the dust mass increases by $\sim0.5$ dex and the corresponding $\delta_{\rm GDR}$ approximately decreases by a factor of 3.

\subsection{Stellar mass and molecular gas fraction}
The logarithmic stellar masses range from 10.3 to 12.1, with a median of 10.8, indicating the majority of our Hot DOGs are massive galaxies. However, \citet{Diaz-santos2021} discovered higher SED-based stellar masses than dynamical masses based on ALMA [C II] observations. They attributed this overestimation to the lower angular resolution of optical/NIR data than that of interferometric [C II] data. We defer a comparative analysis of the stellar and dynamical masses of our sample to a future work. With $M_\star$ derived from SED fitting and molecular gas mass $M_{\rm H_2}$ inferred from CO line observations, we calculated the molecular gas fraction, which is defined as $f_{\rm gas}$ = $M_{\rm H_2}/(M_{\rm H_2}\,+\,M_\star)$, and listed the results in Table \ref{table:properties}. Due to the lack of optical-NIR photometry data for W0248+2705, W0615-5716 and W1248-2154, we cannot obtain SED-based stellar masses $M_\star$ for these three Hot DOGs. To put an upper limit on the molecular gas fraction for these three Hot DOGs, we adopted $M_\star$ = $10^{10.3}\,\rm M_\odot$ which is the minimum $M_\star$ estimated for the other galaxies in our sample. 

Our Hot DOGs have comparable redshift and stellar mass ranges to the SMGs studied in \citet{Miettinen2017} (z $\sim$ 2.3 and log$\,M_\star\,[M_\odot]\,=\, 11.09_{-0.53}^{+0.41}$, respectively). However, the molecular gas fraction of our Hot DOGs ($0.33_{-0.17}^{+0.33}$) is much lower than the SMGs in \citet{Miettinen2017} (0.62$_{-0.23}^{+0.27}$).

\begin{figure}
\centering
\resizebox{1.0\hsize}{!}{\includegraphics{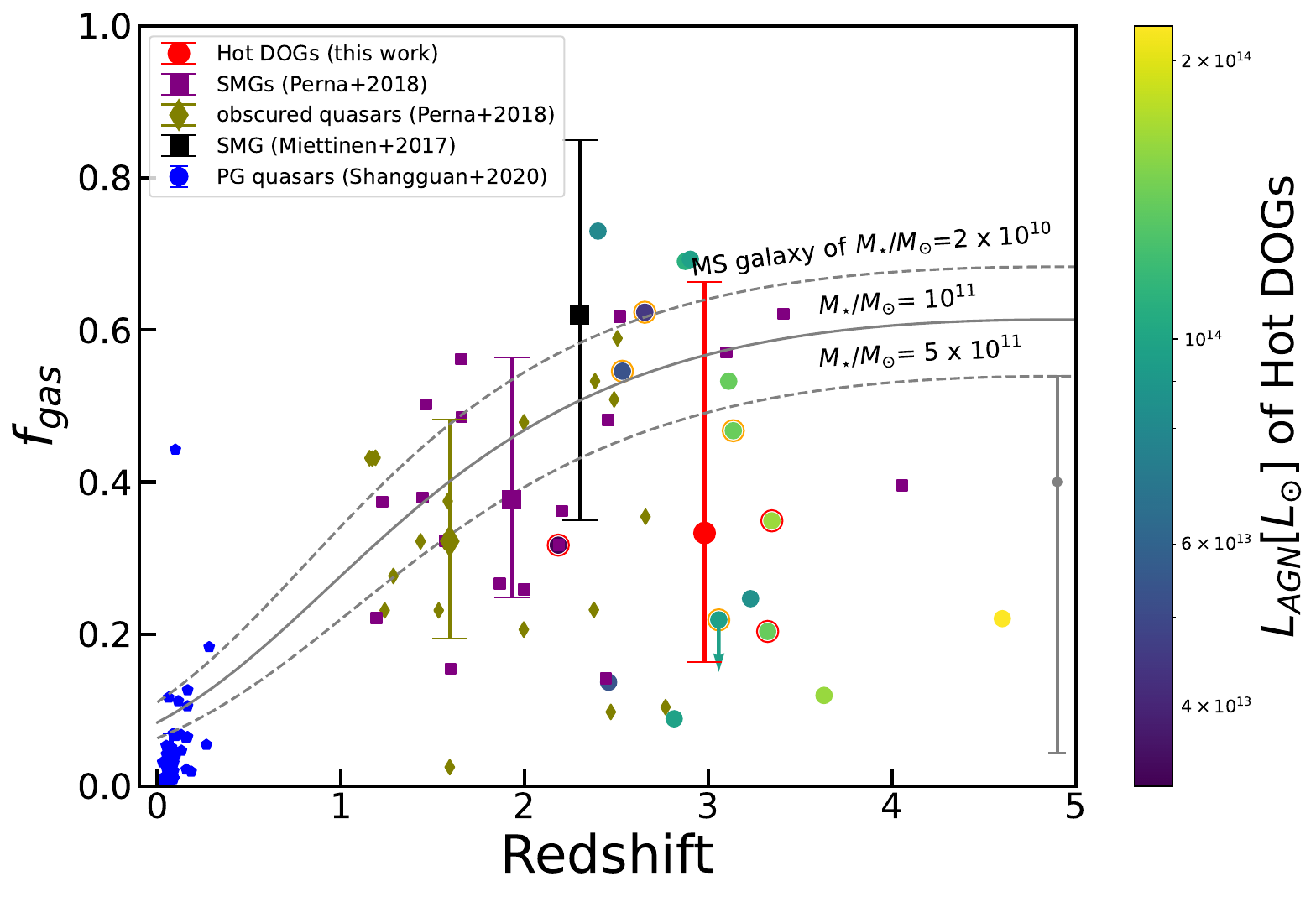}}
\caption{Molecular gas fraction $f_{\rm gas}$ as a function of redshift. The 16 Hot DOGs represented by circle symbols in our sample are color coded according to their AGN bolometric luminosity, and their typical uncertainties are plotted in the bottom-right corner as a gray point with error bars. The three sources that did not have an SED-based stellar mass and were estimated to have a stellar mass of 10$^{10.3}\,M_\odot$ are additionally highlighted with red circles, and four sources whose optical emission are dominated by an AGN component (Section \ref{sec:5.4}) are additionally highlighted with orange circles. The purple square and olive diamond symbols represent the samples of SMGs and obscured quasars, respectively, compiled in \citet{Perna2018}. The black squares represent the sample of SMGs in \citet{Miettinen2017}. Note that \citet{Miettinen2017} have excluded from their sample the SMGs with evidence of AGN, whereas a large fraction of SMGs studied by \citet{Perna2018} have AGN activities. The blue pentagon represents the sample of local PG quasars in \citet{Shangguan2020}.  The median and 16th-84th ranges of our Hot DOGs are shown as a larger filled red circle with an error bar, and the sample median and 16th-84th quantile ranges of the literature samples are shown correspondingly with larger symbols with error bars. The redshift evolutionary trend predicted from the 2-SFM model \citep{Sargent2014} of MS galaxies of $M_\star$ = $10^{11}\,M_\odot$ (typical for our sample) is shown as a solid curve, while MS galaxies of $M_\star$ = $2\,\times\,10^{10}\,M_\odot$ and  $M_\star$ = $5\,\times\,10^{11}\,M_\odot$ are labeled with dashed lines.}
\label{fig:fgas}
\end{figure}

In Figure \ref{fig:fgas}. we show the molecular gas fraction $f_{\rm gas}$ as a function of redshift for our Hot DOGs, as well as literature samples of SMGs, obscured quasars at z $>$ 1 from \citet{Perna2018} and Palomar-Green quasars in the local Universe from \citet{Shangguan2020}. We note that the median uncertainty of $f_{\rm gas}$ of our Hot DOG sample is approximately 60\%, which results from the uncertainties in $M_{\rm H_2}$ and $M_{\star}$ and is shown as gray in the bottom-right corner in Figure \ref{fig:fgas}. The sample of high-redshift SMGs from \cite{Miettinen2017} is also represented in this figure. These literature samples all have $M_\star\,\sim\,10^{11}\,\rm M_\odot$, which is comparable to our Hot DOGs. To demonstrate how the molecular gas fractions of these galaxies compare with MS galaxies, we present the gas fraction evolutionary trend of MS galaxies of $M_\star$ = 10$^{11}\,\rm M_\odot$, $M_\star$ = $2\times10^{10}\,\rm M_\odot$, and $M_\star$ = $5\times10^{11}\,\rm M_\odot$ predicted from the 2-SFM model \citep{Sargent2014}. It is worth noting that a significant proportion of the SMGs compiled in \citet{Perna2018} exhibit AGN activity, while the SMG sample in \citet{Miettinen2017} has excluded SMGs that demonstrate evidence of hosting an AGN. Similar to the PG QSOs from \citet{Shangguan2020} and the SMGs and obscured QSOs from \citet{Perna2018}, the $f_{\rm gas}$ of most of our Hot DOGs is below the relation for MS galaxies with comparable $M_\star$. This is consistent with \citet{Diaz-santos2018}, and \citet{Penney2020}, which concluded that Hot DOGs may have lower cold molecular gas content than ordinary star-forming galaxies. The sample of SMGs lacking AGN exhibits a higher molecular gas fraction compared to MS galaxies. The discrepancy in $f_{\rm gas}$ between active and normal galaxies is likely due to the depletion of cold gas by the AGN feedback. We color coded our Hot DOGs in the figure according to their AGN bolometric luminosity（$L_{\rm AGN}$）derived by integrating the AGN component of the best-fit UV to millimeter SED. Most Hot DOGs with an $f_{\rm gas}$ lower than MS galaxies exhibit a relatively higher $L_{\rm AGN}$, which is consistent with recent findings of a positive correlation between the efficiency of AGN feedback traced by the mass outflow rate and $L_{\rm AGN}$ \citep[see, e.g.][]{Hopkins2016,Fiore2017,Garca2021}. AGN-driven outflows deposit energy and momentum into the surrounding gas and affect the evolution of the host galaxy by heating and ejecting the ISM \citep[e.g.,][]{Weymann1991,Faucher2012,Marasco2020}. Our Hot DOGs stand out from the other samples by their extremely high $L_{\rm AGN}$, which may explain their relatively large deviation from the MS relation.

\subsection{Star formation rate and star formation efficiency}
\begin{figure}
\centering
\resizebox{1.0\hsize}{!}{\includegraphics{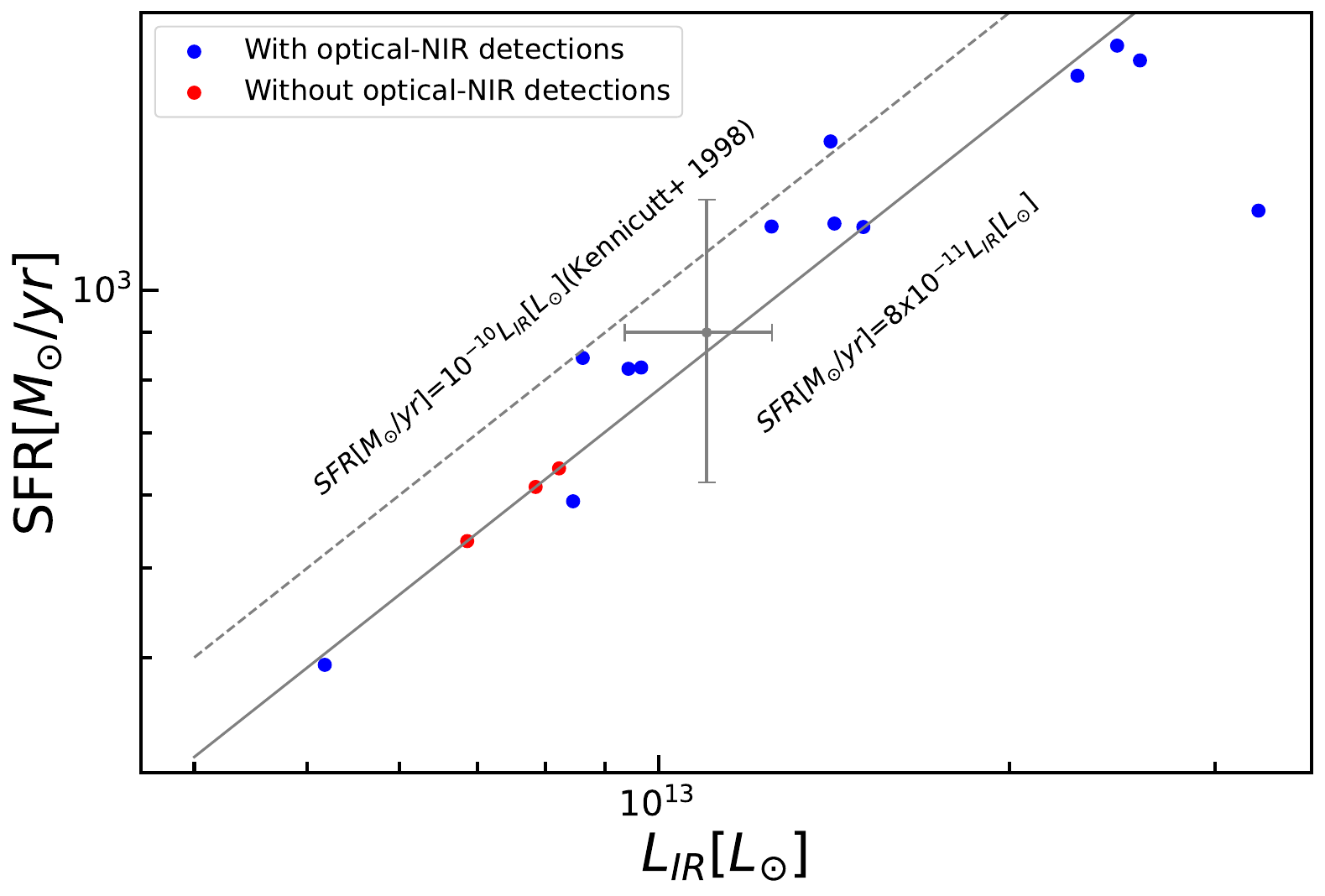}}
\caption{SFR traced by the IR luminosity. The sources with optical-NIR detections are shown as blue circles, and they can be represented by the relation as labeled with a solid curve in the figure, which is about 0.10 dex lower than the \citet{Kennicutt1998} relation calibrated by Chabrier IMF. The SFRs of three Hot DOGs without optical-NIR photometry, namely W0248+2705, W0615-5716, and W1248-2154, are calculated based on this relation. Their typical uncertainties are shown as a gray point with an error bar.}
\label{fig:SFR}
\end{figure}

The SFRs of Hot DOGs were derived by averaging the modeled exponential SFH in the last 100 Myr. The cold dust component is modeled by adding a grey body component, whose energy budget is identical to the attenuated luminosity in the UV-optical band, i.e., the so called energy balance assumption. Figure \ref{fig:SFR} shows that our sample generally follows the relation of SFR $[\rm M_{\odot}/\rm yr]$\,=\,$8\times10^{-11}\,L_{\rm IR}\,[\rm L_{\odot}]$, which is about 0.1 dex lower than the \citet{Kennicutt1998} relation calibrated by the \citet{chabrier2003} IMF. For the three sources without an optical-NIR detection, we estimate their SFRs based on the relation between SFR and $L_{\rm IR}$ calibrated for the remaining 13 Hot DOGs. The SFRs estimated for our Hot DOGs are shown in Table \ref{table:SED fitting results}. By invoking the MS evolutionary model of  \citet{Speagle2014}:
\begin{equation}\label{equ:MS relation}
\begin{aligned}
\log({\rm SFR}/{\rm M}_{\sun}~{\rm yr}^{-1}) &=(0.84-0.026\times \tau_{\rm univ})\log(M_{\star}/{\rm M}_{\sun})\\ 
& \qquad -(6.51-0.11\times \tau_{\rm univ})\, ,
\end{aligned}
\end{equation}

where $\tau_{\rm univ}$ is the age of the Universe in gigayears, we can estimate the offset of our Hot DOGs from the main sequence $\Delta_{\rm MS}$ = $\rm {SFR}\,/\,\rm {SFR}_{\rm MS}$ relation. The high $\Delta_{\rm MS}$ values (Table \ref{table:properties}) demonstrate that most of our Hot DOGs are extreme starburst systems, which could be triggered by gas-rich galaxy mergers \citep[e.g.,][]{Noguchi1986,Mihos1996}. With the molecular gas masses and SFRs, we derived the gas depletion timescale $t_{\rm depl}\,=\,M_{\rm H_2}\,/\,\rm {SFR}$. Similar to obscured quasars \citep[e.g.,][]{Aravena2008, Brusa2018}, our Hot DOGs exhibit a short gas depletion timescale. In contrast, typical starburst galaxies have a $t_{\rm depl}$ value of several 100 hundred megayears \citep[e.g.,][]{Genzel2010,Bothwell2013,Miettinen2017}. Hot DOGs have been discovered to exhibit outflow mass loss rates of several thousand solar masses per year \citep{Finnerty2020}. Considering the powerful AGN outflows, the gas depletion timescale could be shorter. The molecular gas in Hot DOGs may be depleted within several tens of megayears, resulting in the lower molecular gas fraction in Hot DOGs. The few Hot DOGs with a high gas fraction may represent a relatively earlier stage after the AGN activity has been triggered. 

\begin{figure}
\centering
\resizebox{1.0\hsize}{!}{\includegraphics{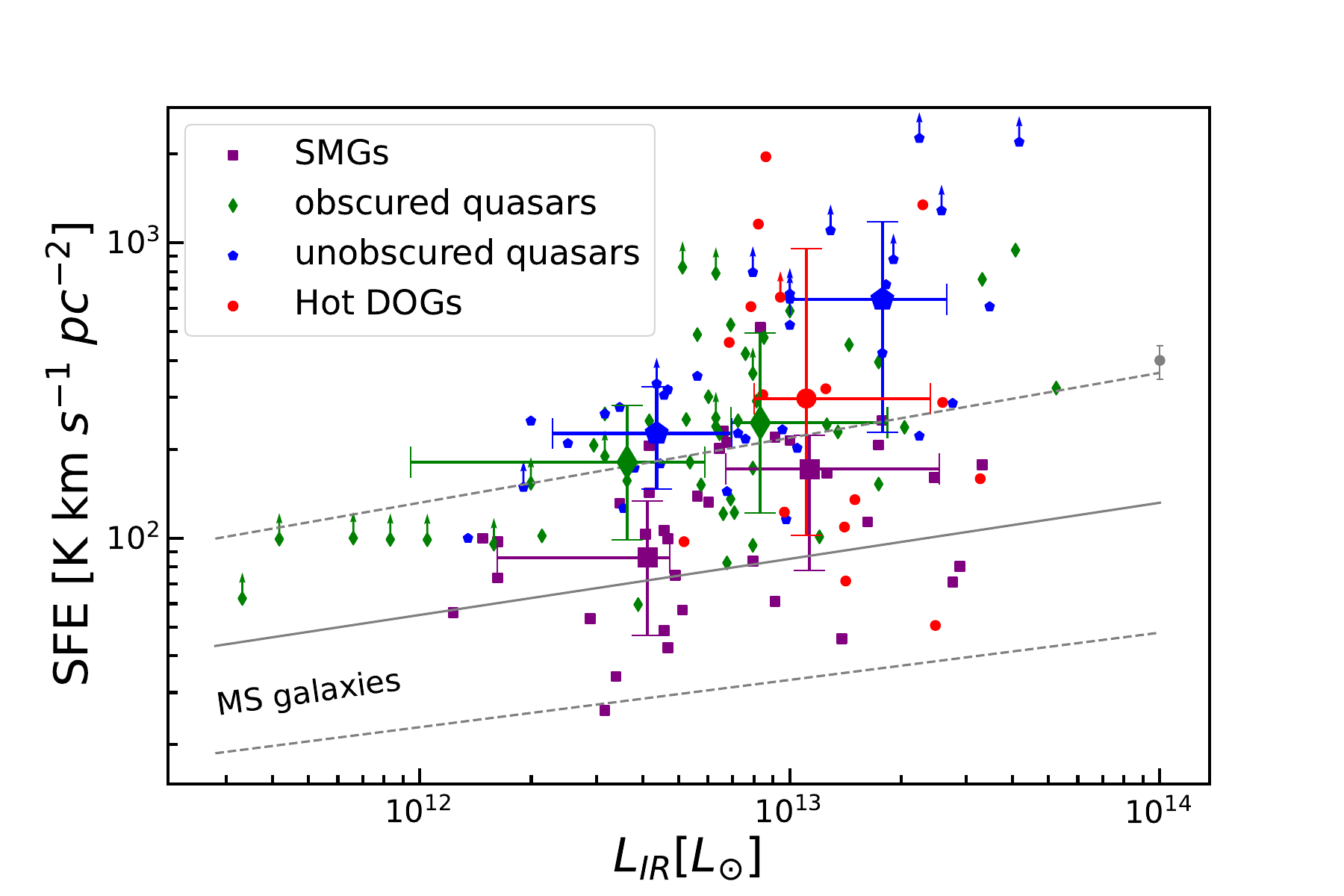}}
\caption{SFE traced by the ratio of IR to CO(1-0) line luminosity as a function of IR luminosity $L_{\rm IR}$. Our sample is labeled with red circles. The typical uncertainties are shown as a gray point with error bars. The purple square, olive diamond and blue pentagons represent the samples of SMGs, obscured quasars, and unobscured quasars, respectively, compiled in \citet{Perna2018}. The upward arrows mark lower limits. Each sample in \citet{Perna2018} has been divided into two bins according to IR luminosity. The median values of the galaxies in various samples mentioned above are shown with larger symbols, overlaid by their 16th-84th quantile ranges as error bars. The solid line shows the best-fit relation for massive MS galaxies, and the 1 $\sigma$ scatter is indicated with two dashed lines \citep{Sargent2014}.}
\label{fig:SFE}
\end{figure}

\begin{figure}
\centering
\resizebox{1.0\hsize}{!}{\includegraphics{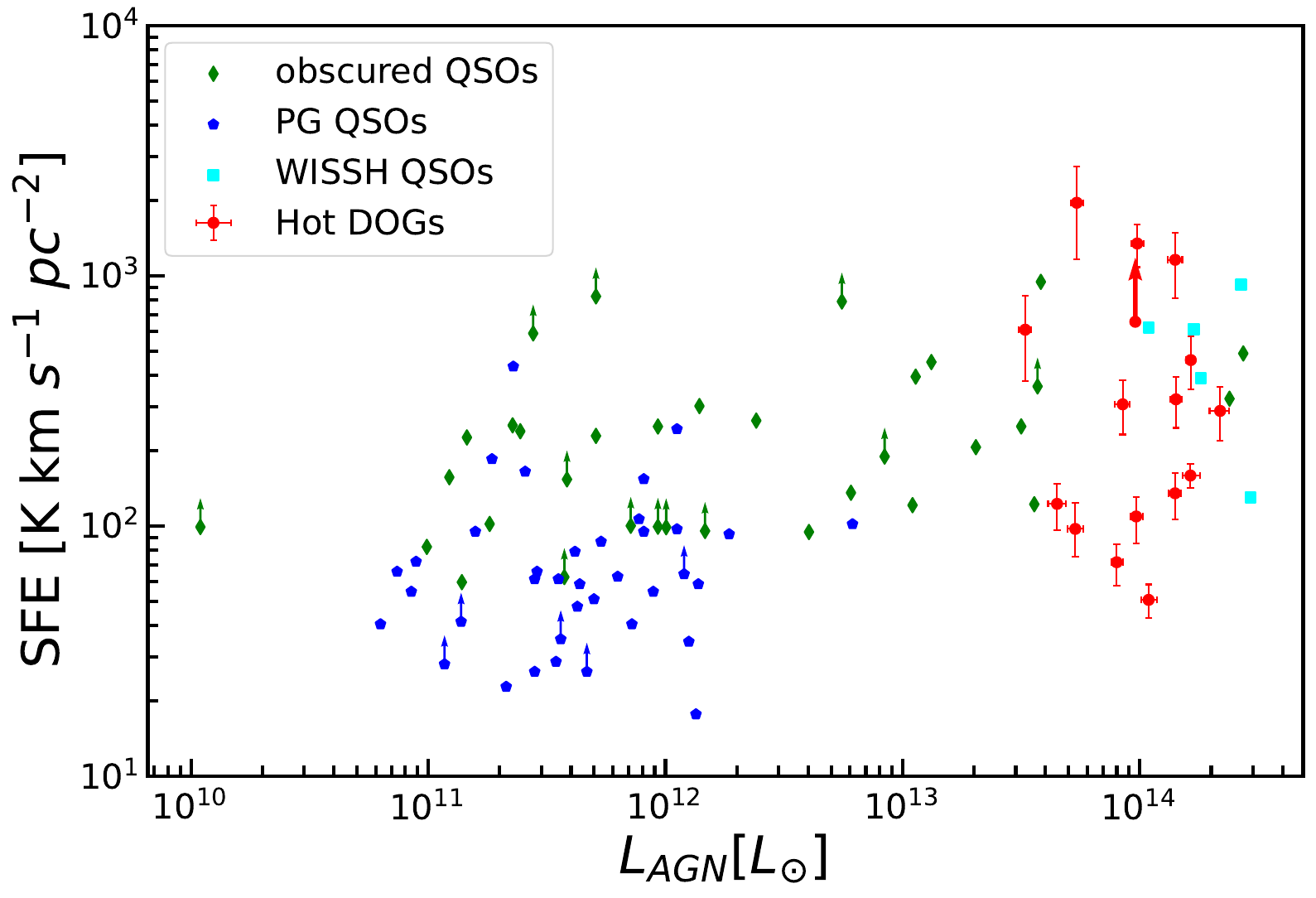}}
\caption{Positive correlation between SFE and AGN bolometric luminosity. Red circles with error bars represent our Hot DOGs. Green diamond symbols represent the sample of obscured quasars in \citet{Perna2018}. The blue pentagons represent the sample of local PG quasars in \citet{Shangguan2020}. Cyan square symbols represent a sample of hyperluminous, type I quasars at z $\sim$ 1-4 in \citet{Bischetti2021}. The downward arrows mark upper limits. }
\label{fig:correlation}
\end{figure}

In Figure \ref{fig:SFE}, we show the SFEs (= SFR / $M_{\rm H_2}$) as a function of IR luminosity $L_{\rm IR}$ for our Hot DOGs, as well as for the compiled samples of SMGs and unobscured and obscured quasars from \citet{Perna2018}. We use the corresponding observables, $L_{\rm IR}$ obtained from the cold dust component of our SED decomposition and $L_{\rm CO(1-0)}^{'}$ acquired through ALMA observations, to calculate the SFE as $L_{\rm IR}$ / $L_{\rm CO(1-0)}^{'}$.  The SFEs exhibit a tight correlation with $L_{\rm IR}$ for both MS galaxies \citep{Sargent2014} and SMGs. Our Hot DOGs, along with obscured and unobscured quasars at z $>$ 1, display high SFEs that are well above the relation for MS galaxies. This is likely due to the rapid depletion of cold gas by AGN feedback and star formation. We plot the relation between SFE and ${L_{\rm AGN}}$ for Hot DOGs and obscured quasars in Figure \ref{fig:correlation}. We also included WISE-SDSS selected hyper-luminous (WISSH) quasars from \citet{Bischetti2021} and PG QSOs from \citet{Shangguan2020}. The AGN bolometric luminosities of the obscured quasars from \citet{Perna2018} is calculated from their X-ray luminosities, assuming a luminosity-dependent bolometric correction from \citet{Duras2020}. Across all four quasar samples, there is a positive correlation between SFE and $L_{\rm AGN}$. The Spearman's rank correlation coefficient $\rho$ = 0.532 (p = 3.55e-8), where p is the probability of the null hypothesis that a correlation does not exist. When excluding the PG QSOs given that they are in local Universe, the $\rho$ deceases to 0.404 with p = 2.21e-3. This correlation can be explained by the enhanced outflow rates in galaxies with high AGN luminosity, and is consistent with their low gas fractions (Figure \ref{fig:fgas}).

\subsection{Influence of AGN contamination to FIR} \label{sec:5.4}
Initially, we fitted our sample with BayeSED by setting the $T_{\rm dust}$ parameter range to default values (10-100K). However, we obtained dust temperatures $\sim60-90K$, which are higher than DOGs and submillimeter galaxies (SMGs), but are consistent with previous studies  employing a similar fitting methodology \citep{Wu2012,Fan2016b,Fan2019}. We refer to this SED fitting as $T_{\rm dust}$-unconstrained fitting and denote the resulting dust temperature and cold dust luminosity as $T_{\rm dust,0}$ and $L_{\rm IR,0}$, listed in Table \ref{table:SED fitting results}. The $L_{\rm IR,0}$ is consistent with those given in Table \ref{tab:sample}. For the $T_{\rm dust}$-unconstrained fitting, the FIR emission is entirely attributed to cold dust heated by massive stars. 

\begin{figure}
\centering
\resizebox{1.0\hsize}{!}{\includegraphics{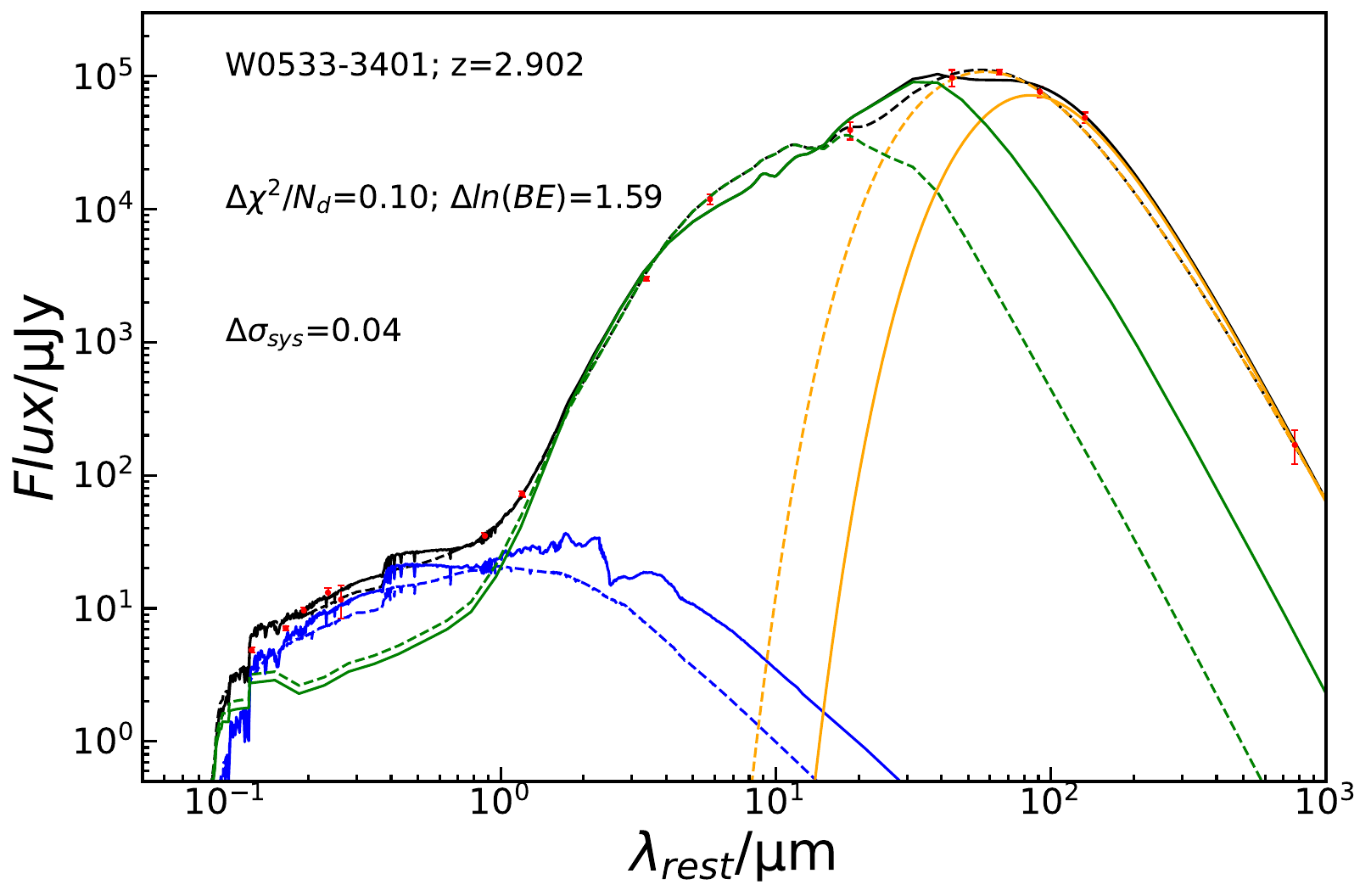}}
\caption{Best-fit model SED comparison between $T_{\rm dust}$-constrained fitting (solid line) and $T_{\rm dust}$-unconstrained fitting (dashed line) for W0533-3401. The source ID, redshift, $\chi^2/N_d$ offset, Bayes factor offset, and the systematical error offset \citep[see][]{HanY2014a} are shown on the panel.  The color of the legend is identical to that of Figure \ref{figure:SED}.}
\label{fig:w0533}
\end{figure}

However, many studies suggested that AGNs could heat dust up to kiloparsec scales, significantly contributing to their FIR emission \citep[see section 7 in][for a recent review]{Netzer2015}. For example, \citet{Schneider2015} conducted radiative transfer modeling and reported that AGN-heated dust contributes from 30$\%$ up to 70\% of the FIR luminosity for a high-redshift quasar host galaxy. \citet{Duras2017} employed the same technique and found 40$\%$ to 60$\%$ quasar contribution to the FIR emission for their sample. \citet{Tsukui2023} determined an AGN contribution of approximately $53\%$ based on image decomposition of spatially resolved ALMA continuum observations. The high cold dust temperature estimated for Hot DOGs may result from contribution by AGN-heated warmer dust, similar to the findings of \citet{Tsukui2023}. 

The hot dust MIR emission from the central torus may be reprocessed to the FIR by optical thick dust in nuclear region \citep[e.g.,][]{Scoville2017,Sokol2023}. Additionally, the FIR contribution of the central AGN may also originate from emission by narrow-line region (NLR) polar dust, which has been proven to be heterogeneous in nearby AGNs \citep[][and references therein]{Netzer2015}. Given the high AGN luminosities of our Hot DOGs, the NLR dust is expected to extend to kiloparsec scales. \citet{2019ApJ...884...11G} improved SED fitting for higher-luminosity AGN by employing two-component models proposed by \citet{Honig2017}, which incorporate a clumpy disk and a polar clumpy wind, in contrast to the CLUMPY model. There is evidence suggesting that the extended polar dust emission is likely associated with AGN-driven dusty outflows \citep[e.g.,][]{Alonso-Herrero2021}. \citet{Lyu2018} utilized their AGN SED libraries, which include an extended polar dust component, to model the SEDs of Hot DOGs. Their results attribute a significant portion of the FIR emission to the polar dust component.

It is crucial to consider accurately the AGN contribution to the FIR emission and to estimate the cold dust luminosity appropriately.  Based on the flux density ratios of 350 $\mu$m and 1.1 mm continuum of several Hot DOGs, \citet{Wu2012,Wu2014} suggested that the cold dust component heated by star formation is not very different from those in starburst galaxies, characterized by temperatures ranging between 30 and 50 K  \citep[e.g.,][]{Magnelli2012}. We constrained the $T_{\rm dust}$ parameter of the graybody component to be within the range 30-50 K and refitted the SEDs in our sample. This fitting approach is denoted as $T_{\rm dust}$-constrained fitting. We show an example in Figure \ref{fig:w0533} which compare the best-fit SEDs obtained via $T_{\rm dust}$-constrained fitting (solid line) and $T_{\rm dust}$-unconstrained fitting (dash line). For the $T_{\rm dust}$-constrained fitting, the AGN emission dominates up to rest-frame 50 $\mu$m, while for the $T_{\rm dust}$-unconstrained fitting, the AGN emission only dominates up to rest-frame 25 $\mu$m. The reduced $\chi^2$ only improve by a value of 0.10 after we apply our prior knowledge of the $T_{\rm dust}$ for W0533-3401. We calculate the Bayes factor, defined as the difference in the Bayesian evidence between the two fitting methods, to be in the range of 0.1-3.6, suggesting no strong evidence in favor of the $T_{\rm dust}$-unconstrained fitting \citep{HanY2014a}. In the earliest torus model, \citet{Pier1993} enlarged the torus to account for the FIR emission by a $\sim$100 pc scale torus. Similarly, for the CLUMPY model of \citet{Nenkova2008a,Nenkova2008b}, the AGN FIR emission can come from an extended torus with a large torus outer radius $R_{\rm out}\,=\,Y\,R_{\rm in}$\citep{Drouart2014}, where $Y$ is the radial extent, one of the six free parameters used to define the CLUMPY model. $R_{\rm in}$ is the inner radius set by the location of the dust at the sublimation temperature, and is computed using the AGN bolometric luminosity $L_{\rm AGN}$ by the equation:
\begin{equation}
    R_{\rm in}\,=\,0.4\,(\frac{L_{\rm AGN}}{10^{45}\, \rm erg\,s^{-1}})^{0.5}\,\rm pc.
\end{equation}
Based on the $T_{\rm dust}$-constrained fitting, we obtain $L_{\rm AGN}\,\sim\,10^{47.6}$ and $Y\,\sim\,65$ for our sample, which give $R_{\rm out}\,\sim\,0.5$ kpc. The $L_{\rm IR}$ is smaller by 0.16-0.47 dex compared to the $L_{\rm IR,0}$ shown in Table \ref{table:SED fitting results}, as a significant proportion of the FIR emission is assigned to the AGN component modeled by CLUMPY model. This is generally consistent with the results of \citet{Diaz-santos2021}. Based on the ALMA dust continuum image measurements, they observed that Hot DOGs do not exhibit a particularly small far-IR size, and the unresolved central AGN component contributes to some extent (from 20\% up to 80\% in the most extreme cases) but does not dominate the FIR emission. Overall, we believe that the $T_{\rm dust}$-constrained fitting method effectively considers the AGN contribution to the FIR. The analysis of the physical properties in this paper is entirely based on the $T_{\rm dust}$-constrained fitting.

\subsection{The Applicability of the CLUMPY Model to Hot DOGs with Excess Blue Light}\label{sec:5.5}
Hot DOGs are usually under heavy dust obscuration, and optically seen as type 2 AGN \citep{Wu2012}. However, \citet{Assef2016} discovered a subpopulation of eight Hot DOGs which exhibit a blue UV-optical SED similar to blue bump by the AGN accretion disk. They named them BHDs. Their rest-frame UV spectra are of typical type 1 quasars, showing broad emission lines \citep{Assef2020}. They confirmed that the UV-optical SEDs of this subpopulation Hot DOGs are due to 1\% scattered light from the highly obscured, hyperluminous AGN into our line of sight, using X-ray and imaging polarization observations \citep{Assef2016,Assef2020,Assef2022}. One Hot DOG in our sample, namely W0220+0137, has been identified as a BHD in \citet{Assef2016,Assef2020}.

The CLUMPY model of \citet{Nenkova2008a,Nenkova2008b} is a geometrical torus model that assumes a certain geometry and dust composition and conducting radiative transfer modeling. The absorption and scattering coefficients given in \citet{Ossenkopf1992} are used, where the UV-optical emission is dominated by the AGN-scattered radiation \citep{Nenkova2008a}. \citet{Ichikawa2015} employed the CLUMPY model to fit the SEDs of type 2 Seyferts with scattered light or a hidden broad-line region (HBLR), which is similar to BHDs, and they identified differences in the modeled torus geometry of HBLRs compared to type 2 Seyferts lacking HBLR signatures.  As shown in Figure \ref{figure:SED}, our SED modeling of W0220+0137 is consistent with the results of \citet{Assef2016,Assef2020}, where the UV-optical SED is dominated by an AGN component. For almost all other Hot DOGs with an optical detection, we found that the UV light is dominated by AGN component. There are three Hot DOGs, namely W0134-2922, W1603+2745 and W2054+0207, whose optical-NIR band also exhibits more AGN emission than stellar emission. These objects are also likely to be compatible with BHDs, albeit to a relatively less extent. For these four BHDs, their stellar mass estimations are more uncertain. For example, \citet{Merloni2010} decomposed the UV to MIR SEDs of 89 type 1 AGNs with two-component SED fitting, and they assigned an upper limit to the stellar mass for galaxies whose contribution of stellar light in the K band is less than 5\%. None of our Hot DOGs exhibit such low stellar contribution, and the cold dust FIR emission from the host galaxy serves as an additional constraint to the stellar component.  We have highlighted these four galaxies in Table \ref{table:SED fitting results} and Figure \ref{fig:fgas}. We also note that W2246-0526 and W2305-0039 have nearly equal contributions from AGN and stellar components in the optical-NIR band.

\subsection{Estimation of the Central Supermassive Black Hole Mass and Growth Rate}
Recent studies of Hot DOGs have consistently revealed that their Eddington ratios are near or above the Eddington limit \citep{Wu2018,Finnerty2020,Jun2020}. \citet{Tsai2018} reported the measurement of $M_{\rm BH}$ of W2246-0526. By using the $L_{\rm AGN}$ of W2246-0526 from our SED fitting, we estimated a super-Eddington ratio $\lambda_{\rm Edd}$ = $L_{\rm AGN}/L_{\rm Edd}$ = 1.7, where $L_{\rm Edd}/L_\odot = 3.28\,\times\,10^4(M_{\rm BH}/{M_\odot})$. Given these studies, we assume an Eddington ratio $\lambda_{\rm Edd}$ = 1.0 for our Hot DOG sample and infer the mass of the SMBH $M_{\rm BH}$ and the black hole to stellar mass ratio $M_{\rm BH}$/$M_\star$. The results are listed in Table \ref{table:properties}.

\begin{figure}
\centering
\resizebox{1.0\hsize}{!}{\includegraphics{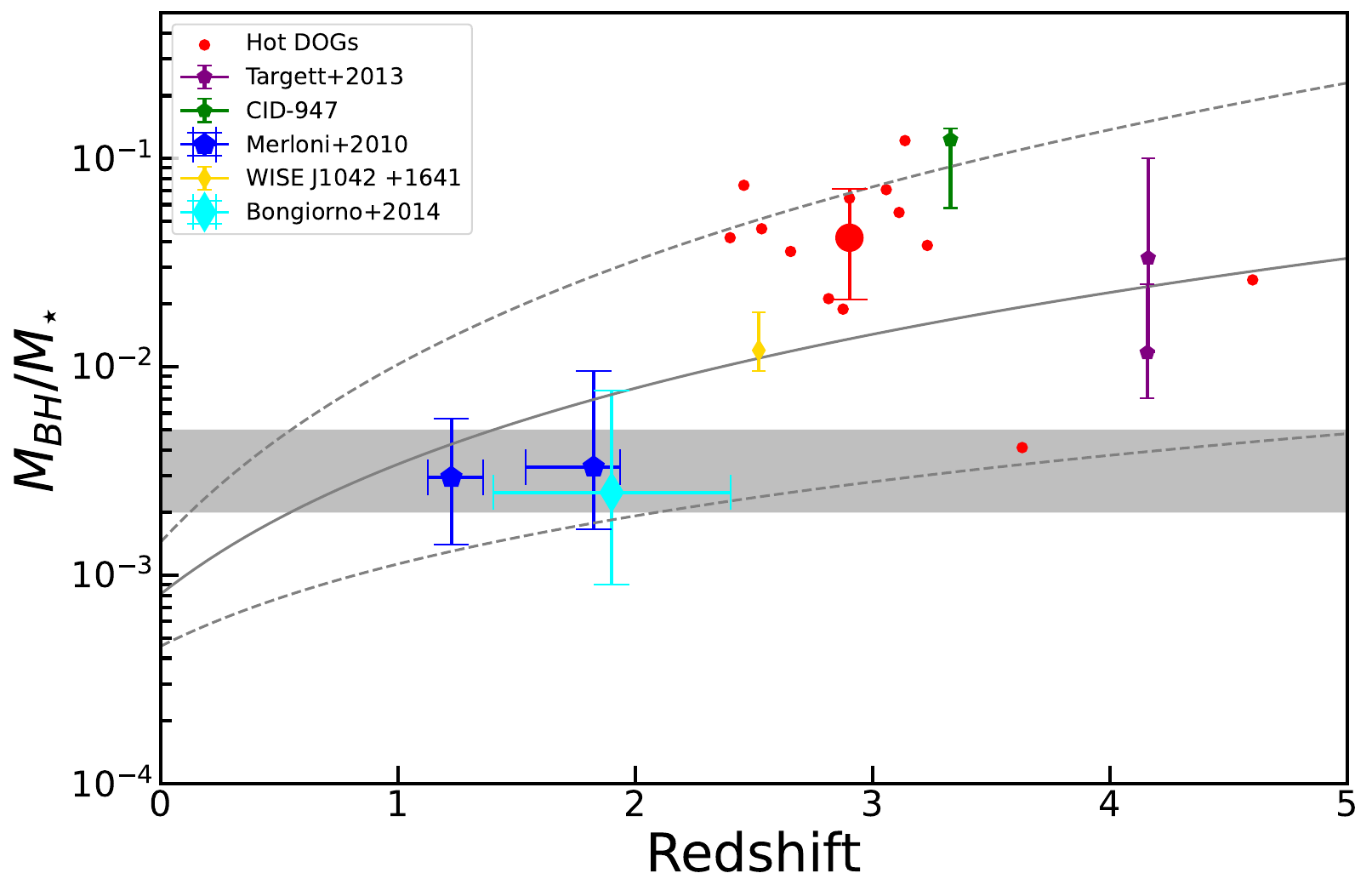}}
\caption{Redshift evolution of black hole to stellar mass ratio ($M_{\rm BH}$/$M_\star$). The $M_{\rm BH}$/$M_\star$ ratios of Hot DOGs are shown as red circles, and the sample median and 16th-84th quantile ranges are shown as a larger symbol with error bars.  Blue triangles represent a sample of 89 moderately luminous broad-line AGN in the redshift range 1 $<$ z $<$ 2.2 \citep{Merloni2010}. A sample of 21 X-ray obscured, red AGNs with moderate luminosity in \citet{Bongiorno2014} is labeled with a cyan diamond symbol. For luminous quasars with AGN bolometric luminosity $>$ $10^{46}\,\rm erg\,\rm s^{-1}$, two luminous SDSS quasars at z $\sim$ 4 \citep{Targett2012} and an extremely red dust-obscured quasar \citep{Toba2016,Matsuoka2018} are denoted with purple filled triangles and gold filled square respectively. The green-filled diamond denotes an X-ray selected luminous unobscured quasar, CID-947 at z $\sim$ 3.3, which has an extremely high black hole to stellar mass ratio $M_{\rm BH}$/$M_\star$ = 1/8 \citep{Trakhtenbrot2015}. The solid line and two dashed lines show the best-fit evolutionary trend of $M_{\rm BH}$/$M_\star$ and 1$\sigma$ errors at z $<$ 2 \citep{McLure2006}. The gray area shows the typical range of $M_{\rm BH}$/$M_\star$ $\sim$ 0.002-0.005 in the local Universe \citep{Kormendy2013}.}
\label{fig:SMBH} 
\end{figure}

In Figure \ref{fig:SMBH}, we plot $M_{\rm BH}$/$M_\star$ as a function of redshift for Hot DOGs and other samples of AGN for different redshift and AGN bolometric luminosity ranges. Similar to other luminous quasars at redshift 2-4 \citep[e.g.,][]{Targett2012,Trakhtenbrot2015,Matsuoka2018}, the inferred $M_{\rm BH}$/$M_\star$ of the Hot DOG sample is about 2 times higher than the evolutionary trend of $M_{\rm BH}$/$M_\star$ seen by \citet{McLure2006}. \citet{Bischetti2021} also revealed an extremely high ratio between SMBH mass and dynamical mass $M_{\rm BH}$/$M_{\rm dyn}$ in WISSH quasars. These WISSH quasars also exhibit a close or super unit Eddington ratio. In contrast, obscured and unobscured AGN with moderate AGN bolometric luminosities at redshift 1-2 \citep[e.g.,][]{Merloni2010,Bongiorno2014} exhibit a relatively low $M_{\rm BH}$/$M_\star$ value. Using the equation  $L_{\rm AGN}\,=\,(\eta {\dot{M}}_{\rm BH}\,c^2)/\,({1-\eta})$ and adopting $\frac{\eta}{1-\eta}$ = 0.1, we derived high black hole growth rates of ${\dot{M}}_{\rm BH}\,\sim\,65_{-29}^{+39}$ $M_\odot\,\rm yr^{-1}$ for our sample. These results suggest rapid black hole growth in Hot DOGs, consistent with \citet{Wu2018}. During this high-accretion phase, the majority of the black hole mass could be assembled within the Salpeter timescale, which is consistent with the gas depletion timescale and the high luminosity state timescale measured in \citet{Tsai2015}.

\section{Summary and Conclusion}\label{section6}
We present a UV to millimeter SED analysis and molecular gas content measurements in a sample of 16  WISE-selected, hyperluminous dust-obscured quasars at z $\sim$ 3. We inferred the physical properties of this sample, such as gas-to-dust ratio, molecular gas fraction, and SFE. This study represents the largest sample to date in which a systematic investigation of the cold gas content in hyperluminous quasars at Cosmic Noon has been conducted. The main results can be summarized as follows:

\begin{enumerate}
    \item Based on ALMA observations of the CO(3-2) or CO(4-3) lines, we have calculated the molecular gas mass $M_{\rm H_2}$ = $3.69_{-2.61}^{+9.92}\,\times\,10^{10}\,M_\odot$ of our sample by adopting $L_{\rm CO(3-2)}^{'}$/$L_{\rm CO(1-0)}^{'}$ = 0.97, $L_{\rm CO(4-3)}^{'}$/$L_{\rm CO(1-0)}^{'}$ = 0.87 and $\alpha_{\rm CO}$ = 0.8 $\rm M_\odot$(K\,km\,s$^{-1}$\,pc$^2$)$^{-1}$. The derived molecular gas masses are higher than the prediction in \citet{Fan2016b}.
    \item We modeled the observed UV to millimeter SEDs of our sample using an updated version of BayeSED. The median values and 16th-84th quartiles of all sources in our sample are given as follows. For the cold dust emission represented by the graybody function, we estimated a typical $T_{\rm dust}\,\sim\,45_{-1}^{+2}\,\rm K$, $\beta\,\sim\,2.5_{-0.2}^{+0.2}$, $L_{\rm IR}\,\sim\,1.1_{-0.3}^{+1.3}\,\times\,10^{13}\,L_\odot$, $M_{\rm dust}$ = $8.0_{-4.8}^{+4.0}\,\times\,10^{7}\,M_\odot$. For the stellar component, we infered typical $M_\star\,\sim5.8_{-2.3}^{+12.3}\,\times\,10^{10}\,M_\odot$ and SFR $\sim\,1008_{-409}^{+597}\,\rm M_\odot\,\rm yr^{-1}$. For the AGN component, we obtained typical $L_{\rm AGN}\,\sim\,9.7_{-4.3}^{+5.8}\,\times\,10^{13}\,\rm L_\odot$.
    \item We estimated the gas-to-dust ratio, finding $\delta_{\rm GDR}\,\sim\,474_{-324}^{+711}$. Most Hot DOGs in our sample exhibited a higher $\delta_{\rm GDR}$ than the typical values for the Milky Way, local star-forming galaxies, and high-redshift SMGs. This discrepancy can potentially be attributed to the intense radiation field and high dust temperatures caused by starbursts and potentially the central AGNs, which may result in an underestimation of $M_{\rm dust}$.
    \item We inferred the molecular gas fraction, finding $f_{\rm gas}\,\sim\,0.33_{-0.17}^{+0.33}$. By comparing our findings with SMGs, obscured quasars from the literature, and the $f_{\rm gas}$-z relation for MS galaxies, we  found that Hot DOGs exhibit a relatively low gas content. This lower gas content is likely attributed to the depletion of gas caused by AGN-driven outflows.
    \item The remarkable offset of our Hot DOGs from the MS, $\Delta_{\rm MS}\,\sim\,6.12^{+5.1}_{-2.9}$ suggests that the majority of our Hot DOGs are extreme starburst systems. The gas depletion timescales, 39$_{-28}^{+85}$ Myr are remarkably short. When comparing the average SFE of $\sim\,297_{-195}^{+659}\,\rm K\,\rm km\,\rm s^{-1}$ $\rm pc^{-2}$ with those of SMGs, obscured and unobscured quasars, as well as MS galaxies, we found that Hot DOGs exhibit higher SFEs similar to optically luminous quasars and obscured quasars, rather than SMGs and MS galaxies. Moreover, we discovered a positive correlation between SFE and AGN bolometric luminosity.
    \item Based on AGN bolometric luminosity, we inferred a typical black hole growth rate $\sim\,65_{-29}^{+39}$ $M_\odot\,\rm yr^{-1}$ and a typical black hole mass $\sim\,3.0_{-1.3}^{+1.8}\,\times\,10^{9}\,M_\odot$ by adopting $\frac{\eta}{1-\eta}$ = 0.1 and $\lambda_{\rm Edd}$ = 1.0, respectively. These results suggest that the majority of the black hole mass can be assembled within a Salpeter timescale, which is consistent with the gas depletion timescale and the high luminosity state timescale of Hot DOGs suggested by \citet{Tsai2015}. The observed black hole to stellar mass ratio $\sim\,0.042_{-0.026}^{+0.029}$ is similar to other high-redshift luminous quasars.
\end{enumerate}

We conclude that our results are consistent with the scenario that our sample represents a phase when both star formation and AGN activity are at their peak, leading to rapid depletion of gas and dust, ultimately  transiting the galaxies toward unobscured quasars.

%
%______________________________________________________________

% % % % % % % % % % % % % % % %
%   DISCUSSION 
% % % % % % % % % % % % % % % %

\section*{Acknowledgements}

We thank the anonymous referee for constructive comments and suggestions. This work is supported by National Key Research and Development Program of China (2023YFA1608100) and the Strategic Priority Research Program of Chinese Academy of Sciences, grant No. XDB 41000000. L.F. gratefully acknowledges the support of the National Natural Science Foundation of China (NSFC; grant Nos. 12173037, 12233008), the CAS Project for Young Scientists in Basic Research (No. YSBR-092), the China Manned Space Project (NO. CMS-CSST-2021-A04 and NO. CMS-CSST-2021-A06), the Fundamental Research Funds for the Central Universities (WK3440000006), Cyrus Chun Ying Tang Foundations and the 111 Project for ``Observational and Theoretical Research on Dark Matter and Dark Energy'' (B23042). Y.K. acknowledges the NSFC (grant Nos. 11773063 and 12288102), the ``Light of West China'' Program of Chinese Academy of Sciences, and the Yunnan Ten Thousand Talents Plan Young \& Elite Talents Project. We thank the staff of the Nordic ALMA Regional Center node for their support and helpful discussions. K.K. acknowledges support from the Swedish Research Council and the Knut and Alice Wallenberg Foundation. This paper makes use of the following ALMA data: ADS/JAO.ALMA\#2017.1.00441.S and ADS/JAO.ALMA\#2017.1.00358.S. ALMA is a partnership of ESO (representing its member states), NSF (USA) and NINS (Japan), together with NRC (Canada) and NSC and ASIAA (Taiwan) and KASI (Republic of Korea), in cooperation with the Republic of Chile. The Joint ALMA Observatory is operated by ESO, AUI/NRAO and NAOJ.

Based on observations made with ESO Telescopes at the La Silla Paranal Observatory under program IDs 177.A-3016, 177.A-3017, 177.A-3018 and 179.A-2004, and on data products produced by the KiDS consortium. The KiDS production team acknowledges support from: Deutsche Forschungsgemeinschaft, ERC, NOVA and NWO-M grants; Target; the University of Padova, and the University Federico II (Naples).

\vspace{5mm}
\facilities{ALMA, WISE, Herschel (PACS, SPIRE), CTIO (DECam), Paranal (OmegaCam, VISTA), Bok (90prime), SDSS}

\end{CJK*}

\end{document}